\documentclass[a4paper,aps,prd,preprintnumbers,showpacs,superscriptaddress,nofootinbib,amsmath,amssymb,twocolumn]{revtex4-2}

\usepackage{tabularray}
\usepackage{graphicx}
\usepackage{bm}
\usepackage{soul}
\usepackage{placeins}
\usepackage{xspace}
\usepackage{bm,amsmath,amssymb,amsfonts,bbold}
\usepackage{multirow}
\usepackage{comment}
\usepackage{hyperref}
\hypersetup{
    colorlinks=true,
    linkcolor=blue,
    citecolor=blue,
    filecolor=blue,      
    urlcolor=blue,
}
\usepackage{nicefrac}
\usepackage{cleveref}
\usepackage{color}

\newcommand{\eq}[1]{\begin{align} #1 \end{align}}

\usepackage[english]{babel}
\usepackage{verbatim}
\usepackage{makecell}

\usepackage{ amssymb }
\usepackage{bbold,dsfont}

\begin{document}
\title{Quasinormal Modes and Hawking Radiation of Black Holes with Primary Scalar Hair}

\author{Roman A. Konoplya}
\email{roman.konoplya@gmail.com}
\affiliation{Research Centre for Theoretical Physics and Astrophysics, Institute of Physics, Silesian University in Opava, Bezru\v{c}ovo n\'am\v{e}st\'i 13, CZ-74601 Opava, Czech Republic}

\author{Oleksandr S. Stashko}
\email{alexander.stashko@gmail.com}
\affiliation{Institute of Cosmology, Department of Physics and Astronomy, Tufts University, Medford, Massachusetts 02155, USA}

\author{Zden\v{e}k Stuchl\'ik}
\email{zdenek.stuchlik@physics.slu.cz}
\affiliation{Institute of Physics in Opava, Silesian University in Opava, Bezru\v{c}ovo n\'am\v{e}st\'i 1150/13, 746 01 Opava, Czech Republic}

\begin{abstract}
Recently, a new family of asymptotically flat black-hole solutions endowed with primary scalar hair has been discovered in beyond-Horndeski gravity.  We  study in detail the quasinormal modes spectra,  graybody factors, and Hawking radiation of this class of black holes. We demonstrate that presence of primary scalar hair leaves characteristic imprints on the ringdown properties, shifts the quasinormal frequencies, inducing overtone rearrangements, and  rise of echoes. While the fundamental modes associated with the light-ring are affected moderately, higher overtones are highly sensitive to the small near-horizon deformation produced by scalar field. In certain parameter regimes, the graybody factors exhibit resonant-tunneling behavior, which leads to an oscillatory frequency dependence of the Hawking emission rate. Thus, both black-hole spectroscopy and Hawking radiation may provide complementary and distinctive probes of the beyond-Horndeski gravity. Additionally we demonstrate that the corresponding naked singularites  are quantum mechanically singular and  do not admit a well defined dynamics.
\end{abstract}

\maketitle
\setlength{\parindent}{0pt}
\setlength{\parskip}{1\baselineskip}

\section{Introduction}
The relaxation of a black hole after an external perturbation is governed by a discrete set of complex frequencies determined by the spacetime geometry and by the boundary conditions imposed at the horizon and at infinity. These quasinormal modes (QNMs) are not normal modes in the usual conservative sense: their imaginary parts describe the irreversible loss of energy through the event horizon and by radiation to infinity, and their real parts set the characteristic oscillation scales of the ringdown signal. For this reason, QNMs provide a useful bridge between the mathematical theory of black-hole perturbations, stability analysis, and gravitational spectroscopy~\cite{Chandrasekhar:1983,Kokkotas:1999bd,Berti:2009kk,Konoplya:2011qq,Nollert:1999ji}. The properties of the QNM spectrum and its sensitivity to near-horizon geometry and light-ring physics make it a natural probe of black holes beyond general relativity and of the no-hair hypothesis~\cite{Arun_2022,Berti:2025hly}. Compared with the fundamental mode, which is mainly associated with light-ring physics, overtones are of special interest because they are strongly sensitive to deformations of the near-horizon geometry and can therefore provide additional spectroscopy tests~\cite{Berti:2025hly}. Including overtones in the description of the ringdown phase allows one to reproduce the early-ringdown waveform already close to the peak of the strain, whereas the fundamental mode alone becomes accurate only at later times in the ringdown~\cite{Giesler:2019uxc,Giesler:2024onl}.

Scalar-tensor theories are especially interesting in this respect, since they introduce a natural additional degree of freedom through a spin-0 field. The most general scalar-tensor theory with second-order field equations is Horndeski theory \cite{Horndeski:1974wa}. Its healthy higher-derivative extensions include beyond-Horndeski and DHOST theories \cite{Langlois:2015cwa,BenAchour:2016fzp,Kobayashi:2019hrl}, in which the degeneracy of the Lagrangian prevents the propagation of the would-be Ostrogradsky ghost modes. 

The no-hair hypothesis states that stationary black holes in General Relativity are characterized only by their mass, electric charge, and angular momentum, and are therefore described by the Schwarzschild, Reissner-Nordström, Kerr, or Kerr-Newman solutions, depending on the corresponding set of conserved charges. In scalar-tensor theories, black-hole and compact-object solutions may evade this hypothesis by supporting nontrivial scalar-field profiles. Such solutions may possess secondary scalar hair, for which the scalar field is nontrivial but the corresponding scalar charge is not an independent global charge. Instead, it is fixed by the other parameters of the solution and by the coupling constants of the theory \cite{Herdeiro:2015waa}.

In contrast, primary scalar hair corresponds to the presence of an independent scalar charge and therefore introduces new parameters in compact-object solutions beyond the standard General Relativity families. Recently, families of black holes and compact objects with primary scalar hair have been found in beyond-Horndeski theories \cite{Bakopoulos:2022nqd,Baake:2023zsq,Bakopoulos:2023sdm}. These solutions include black holes, regular black holes, solitons, and exotic compact objects. Additional new  related solutions were also obtained by applying disformal transformations to known primary-hair black holes, thereby mapping the seed configurations to solutions of other DHOST theories \cite{Charmousis:2025xug}. The observational properties of these spacetimes have been studied in \cite{Huang:2026ubc,Erices:2024lci,Nozari:2026wjo}. However, their gravitational-wave phenomenology remains less developed. In particular, gravitational axial perturbations were studied in \cite{Charmousis:2025xug}, while Hawking radiation and quasinormal mode spectra were considered only for some particular cases  \cite{Antoniou:2025bvg}.

In this work we focus on the asymptotically flat black-hole family with primary scalar hair obtained in the shift-symmetric and parity-symmetric sector of beyond-Horndeski gravity~\cite{Bakopoulos:2022nqd} and generalized in \cite{Baake:2023zsq,Charmousis:2025xug}. For the special case considered below, the metric is described by a two-parameter family of solutions, specified by the mass parameter $M$ and the scalar-hair parameter $\xi$.

In our analysis, we restrict ourselves to the propagation of test fields on this fixed background. We study in detail the quasi-normal spectra and the associated scattering problem, compute the graybody factors, and calculate the corresponding energy-emission rates. This approximation is motivated by the fact that, in representative four-dimensional cases, Standard Model fields dominate Hawking radiation and account for about $98\%$--$99\%$ of the radiated power~\cite{Page:1976df}. Consequently, test scalar, electromagnetic, and Dirac fields are sufficient to capture the qualitative dependence of the Hawking emission on the black-hole parameters, although a separate analysis of gravitational perturbations is required for a complete stability and spectroscopy study.

The paper is organized as follows. In Sec.~II we provide basic relation about the parity-symmetric beyond-Horndeski solutions. In Sec.~III we present the test-field equations, boundary conditions, numerical strategy for QNMs, and scattering setup for graybody factors. Sec.~IV contains the numerical results for the spectra, transmission coefficients, and emission rates. We conclude with a summary of the main physical trends.

\section{Parity-symmetric beyond-Horndeski black hole}
The action of the beyond-Horndeski theory has the form \cite{Gleyzes_2015}

\begin{equation}
S = \int d^4x\,\sqrt{-g}\,\bigl(\mathcal{L}_2 + \mathcal{L}_4 + \mathcal{L}^{\rm bH}_4\bigr),
\end{equation}
with
\begin{align}
\mathcal{L}_2 &= G_2(X), \\
\mathcal{L}_4 &= G_4(X)R + G_{4X}(X)\Bigl[(\Box\phi)^2 - \phi_{\mu\nu}\phi^{\mu\nu}\Bigr], \\
\mathcal{L}^{\rm bH}_4 &= F_4(X)\,\varepsilon^{\mu\nu\rho}{}_{\sigma}\,\varepsilon^{\mu'\nu'\rho'\sigma}\,\phi_{\mu}\phi_{\mu'}\phi_{\nu\nu'}\phi_{\rho\rho'}.
\end{align}
where we use the shorthand notation $\phi_\mu=\partial_{\mu}\phi$, $\phi_{\mu\nu}=\nabla_{\mu}\nabla_{\nu}\phi$, $G_{4X}=dG_4/dX$, while  $X=-(\partial \phi)^2/2$  is the canonical kinetic term. The scalar field is assumed to possess shift symmetry ($\phi\to\phi+\mathrm{const}$) and parity symmetry ($\phi\to-\phi$) respectively.

For the exact hairy solution discussed below, the remaining free functions are chosen as
\begin{align}
G_2(X) &= -\frac{2\alpha}{\lambda^2} X^p, \\
G_4(X) &= 1-\alpha X^p, \\
F_4(X) &= \frac{\alpha}{4}(2p-1)X^{p-2},
\end{align}
where $\alpha$ is a dimensional constant that controls the strength of the higher-derivative scalar coupling, $\lambda$ sets the length scale. 

With this choice, the field equations admit a static, spherically symmetric, asymptotically flat branch with line element
\begin{equation}\label{eq:line-element}
ds^2 = -f(r)dt^2 + \frac{dr^2}{f(r)} + r^2(d\theta^2 + \sin^2\theta\,d\varphi^2),
\end{equation}
and metric function
\begin{equation}\label{eq:metric-function}
f(r) = 1  - \frac{2M_0}{r}-\frac{2\xi r^2}{3\lambda^2}{}_2F_1\left(\frac{3}{2},p,\frac{5}{2},-\frac{r^2}{\lambda^2}\right),
\end{equation}
where
\begin{equation}
M_0=M-\frac{\sqrt{\pi } \lambda  \xi  \Gamma
   \left(p-\frac{3}{2}\right)}{4 \Gamma (p)},~\xi=\alpha(2p-1)\frac{q^{2p}}{2^p},
\end{equation}
and for scalar field, we have
with  that 
\eq{
\phi(t,r)=\psi(r)+qt.
}

Here ${}_2F_1(a,b,c,z)$ denotes the standard Gaussian hypergeometric function, while
$\Gamma(z)$ is the Euler gamma function,
$M$ denotes the mass of the configuration, and $q$ plays the role of the primary scalar charge. The special values $p=1/2$ and $p=3/2$ require separate treatment. 

\begin{figure}
    \centering
    \includegraphics[width=0.75\linewidth]{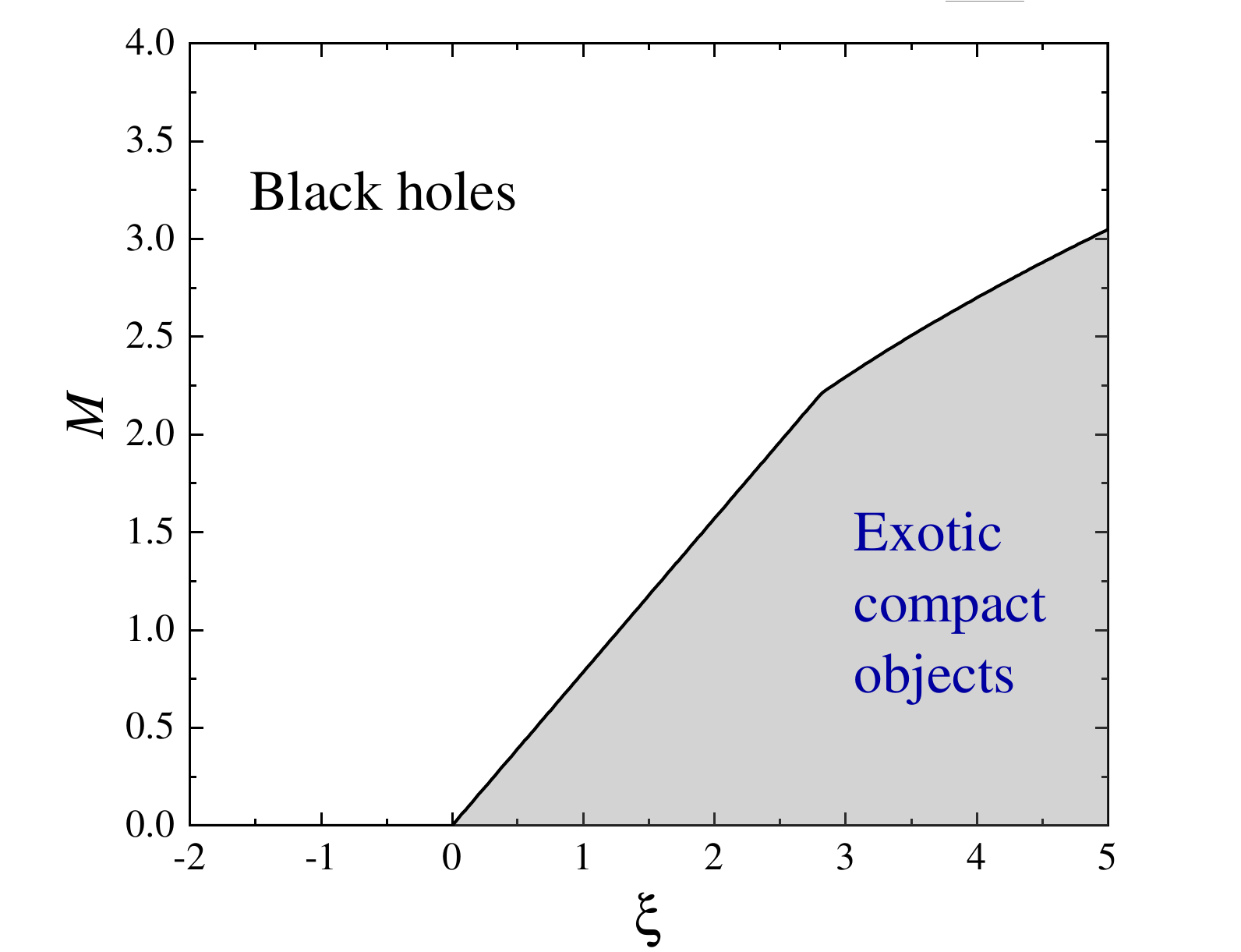}
    \caption{Classification of the solution branches in the $(M,\xi)$ parameter plane for the  (\ref{eq:metric_function}) parity-symmetric beyond-Horndeski geometry. The plot indicates the domains corresponding to black-hole configurations and various exotic compact objects.}
    \label{fig:placeholder}
\end{figure}

We restrict our analysis to the special case $p=2$ that was originally obtained in \cite{Bakopoulos_2024}. We also measure all lengths in units of $\lambda$ and, after the rescaling $M/\lambda\to M$ and $r/\lambda\to r$, keep the same symbols for the dimensionless variables. The metric function then takes the simple form
\begin{equation}
\label{eq:metric_function}
    f(r)=1-\frac{4M-\pi \xi}{2r}-\frac{\xi\arctan(r)}{r}+\frac{\xi}{r^2+1}
\end{equation}
A detailed discussion of the global structure of this family can be found in Refs.~\cite{Bakopoulos_2024,Baake:2023zsq,Charmousis:2025xug}. For our purposes it is important that the geometry is controlled by the two parameters $(M,\xi)$.
Depending on their values, the same analytic expression describes black-hole branches with different central behavior, solitonic configurations, and naked singularities. In the black-hole sector the number of horizons is not fixed a priori: one, two, or three horizons may occur, and for some ranges of $M$ and $\xi$ the outer horizon relevant for wave propagation changes branch. The parameter domains are illustrated in Fig.~\ref{fig:placeholder}.
\section{Test fields}
\label{sec:TestFields}
We consider massless scalar, electromagnetic, and Dirac test fields propagating in the background~\eqref{eq:line-element}. After separation of variables (see, for instance  \cite{Carter:1968ks,Konoplya:2018arm}), the perturbation equations reduce to
\begin{equation}\label{eq:generic-master}
\left[\frac{\partial^2}{\partial t^2}
-\frac{\partial^2}{\partial r_*^2}
+V^{(i)}_{\rm eff}(r)\right]\Psi_i(t,r)=0,
\end{equation}
where $i=(S,V,D)$ corresponds to the scalar, electromagnetic, and Dirac fields. The tortoise coordinate is defined in the standard way,
\begin{equation}
\frac{dr_*}{dr}=\frac{1}{f(r)}.
\end{equation}
The effective potentials are
\eq{\label{eq:VeffSF}
V^{(S)}_{\rm eff}(r)&=f(r)\left[\frac{\ell(\ell+1)}{r^2}+\frac{f'(r)}{r}\right],~\ell\geq 0
}
\eq{\label{eq:VeffVF}
V^{(V)}_{\rm eff}(r)&=f(r)\frac{\ell(\ell+1)}{r^2},
~ \ell\geq1
}
while for the Dirac field one obtains two isospectral potentials,
\begin{equation}\label{eq:dirac-potentials}
V^{(D)\pm}_{\rm eff}(r)=W^2(r)\pm\frac{dW}{dr_*},
~
W(r)=\frac{\lambda\sqrt{f(r)}}{r},
\end{equation}
where $\lambda\geq1$.

Separating the time dependence as
\begin{equation}
\Psi_i(t,r)=e^{-i\omega t}\psi_i(r),
\end{equation}
the quasinormal-mode boundary conditions are
\begin{equation}
\begin{aligned}
\psi_i&\sim e^{-i\omega r_*},
&& r_*\to -\infty,
\\
\psi_i&\sim e^{+i\omega r_*},
&& r_*\to +\infty .
\end{aligned}
\end{equation}
We numerically solve the corresponding boundary-value problem using a Chebyshev pseudospectral method~\cite{Boyd:2001,Trefethen:2000}. Singular contributions at the horizon and at spatial infinity can be removed by factoring out the asymptotic behavior,
\eq{
\psi_i(r)= (r-r_h)^{-i\omega/{f'(r_h)}}  r^{2i M \omega} e^{i\omega r}\psi_i(r),
}
where $\psi_i$ is regular at the horizon and at spatial infinity. Then we introduce the compactified coordinate
\eq{
r=\frac{r_h}{1-u},~u\in(0,1).
}
The coordinate values $u$ are distributed on the $N+1$ Chebyshev--Lobatto nodes
\eq{
u_j=\frac12\left(1-\cos\frac{\pi j}{N}\right),~j=0,\ldots,N.
}
The discretization transforms Eq.~\eqref{eq:generic-master} into
\begin{equation}\label{eq:discr_eq}
\mathbb{M}(\omega)\,\vec{\psi}=0,
\end{equation}
where $\vec{\psi}$ collects the values of the regularized wave function
at the Chebyshev--Lobatto collocation nodes, while
$M(\omega)$ is the $(N+1)\times(N+1)$ matrix representation of the radial wave operator obtained through spectral discretization, quadratic in $\omega$. 
After linearizing and solving the corresponding  eigenvalue problem we obtain the quasinormal spectrum. To avoid spurious nonphysical eigenvalues, we perform the calculations on multiple grids with different node distributions. We use $N=100$--$500$ nodes.

We complement and cross-check our calculations in the frequency domain via time-domain integration with consequent extraction of frequencies via the Prony method.  For this purpose, the wave equation (\ref{eq:generic-master}) is rewritten  as a first-order system on a uniform grid in the tortoise coordinate $r_*$. We approximate spatial derivatives with fourth-order finite differences and evolve the resulting semi-discrete system using the standard fourth-order Runge-Kutta method. The initial data are chosen to be a Gaussian wave packet in form $\Psi=e^{-(r_*-r{}_0)^2/2}$.

To determine quasinormal frequencies from the time-domain profiles, we employed the Prony method \cite{Prony:1795}, which approximates the signal by a superposition of damped exponentials,
\begin{equation}
\Psi(t)\approx \sum_{k=1}^{p} C_k e^{-i\omega_k t},
\end{equation}
where $C_k$ are complex amplitudes and $\omega_k$ are the quasinormal frequencies.

For scattering problems, $\omega$ is real and
\begin{equation}
\begin{aligned}
\psi_i&\sim T_i(\omega)e^{-i\omega r_*},
&& r_*\to -\infty,
\\
\psi_i&\sim e^{-i\omega r_*}+R_i(\omega)e^{i\omega r_*},
&& r_*\to +\infty,
\end{aligned}
\end{equation}
where $R_i(\omega)$ and $T_i(\omega)$ are the reflection and transmission amplitudes. The graybody factor is defined as
\begin{equation}
\Gamma_{i\ell}(\omega)=|T_i(\omega)|^2
=1-|R_i(\omega)|^2.
\end{equation}
Although the scattering problem can also be solved by pseudospectral methods  \cite{boyd1990chebyshev,2014BrJPh..44..128C}, the direct integration is technically simpler. We expand $\psi(r)$ near the horizon and use first $10$ terms of the expansion to as initial conditions for numerical integration from the vicinity of the horizon up to some point $r_{inf}\gg r_h$, where the solution is matched to the corresponding asymptotic expansion at spatial infinity. After extracting the transmission coefficients, we can  calculate the energy emission spectrum as
\begin{equation}
\label{eq:emmision_spectra}
\frac{d^2E_i}{dt\,d\omega}
=\frac{1}{2\pi}\sum_{\ell}
\frac{ \omega\,n_i\Gamma_{i\ell}(\omega)}{
\exp(\omega/T_h)\pm1},
\end{equation}
where the minus sign corresponds to bosons and the plus sign to fermions, $T_h=f'(r_h)/4\pi$ is the Hawking temperature, and $n_i$ is the degeneracy factor, $n_S=2\ell+1$, $n_V=2(2\ell+1)$, and $n_D=4\lambda$, respectively.
The emission spectrum (\ref{eq:emmision_spectra}) is evaluated in the canonical ensemble, where the black-hole parameters and the corresponding Hawking temperature $T_h$ are kept fixed and the back-reaction of the emitted radiation is neglected.

\begin{figure*}
\includegraphics[width=0.33\linewidth]{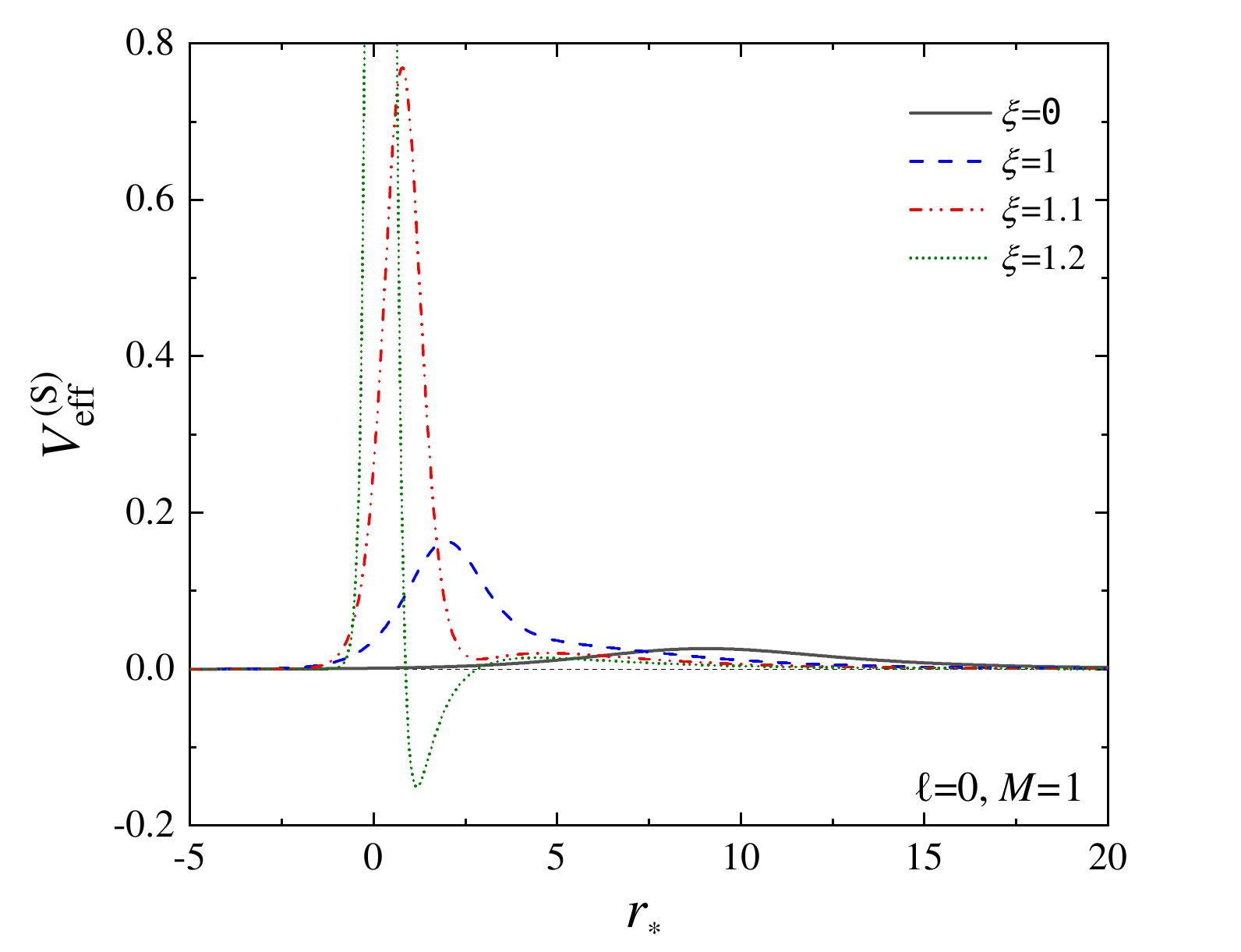}%
\includegraphics[width=0.33\linewidth]{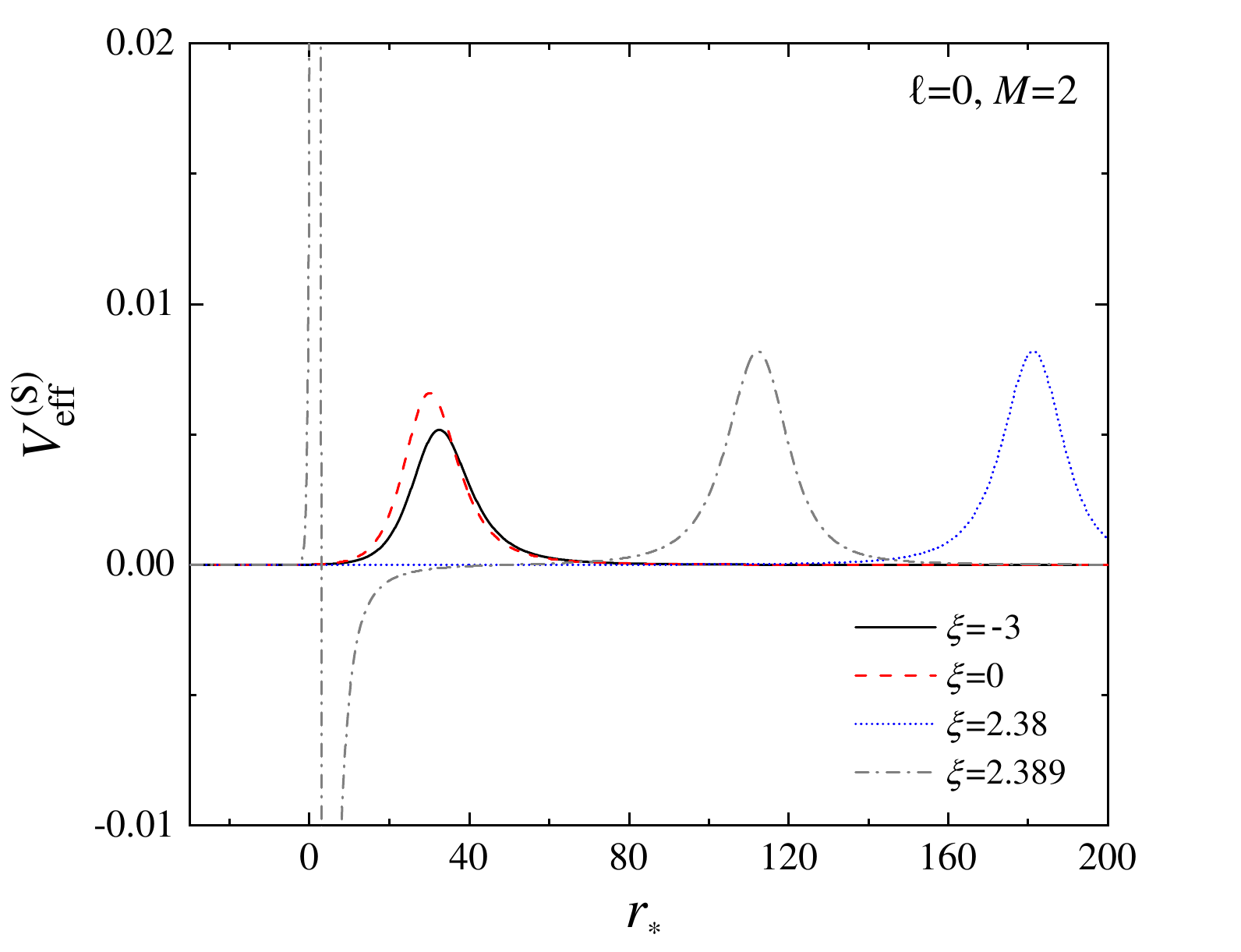}%
\includegraphics[width=0.33\linewidth]{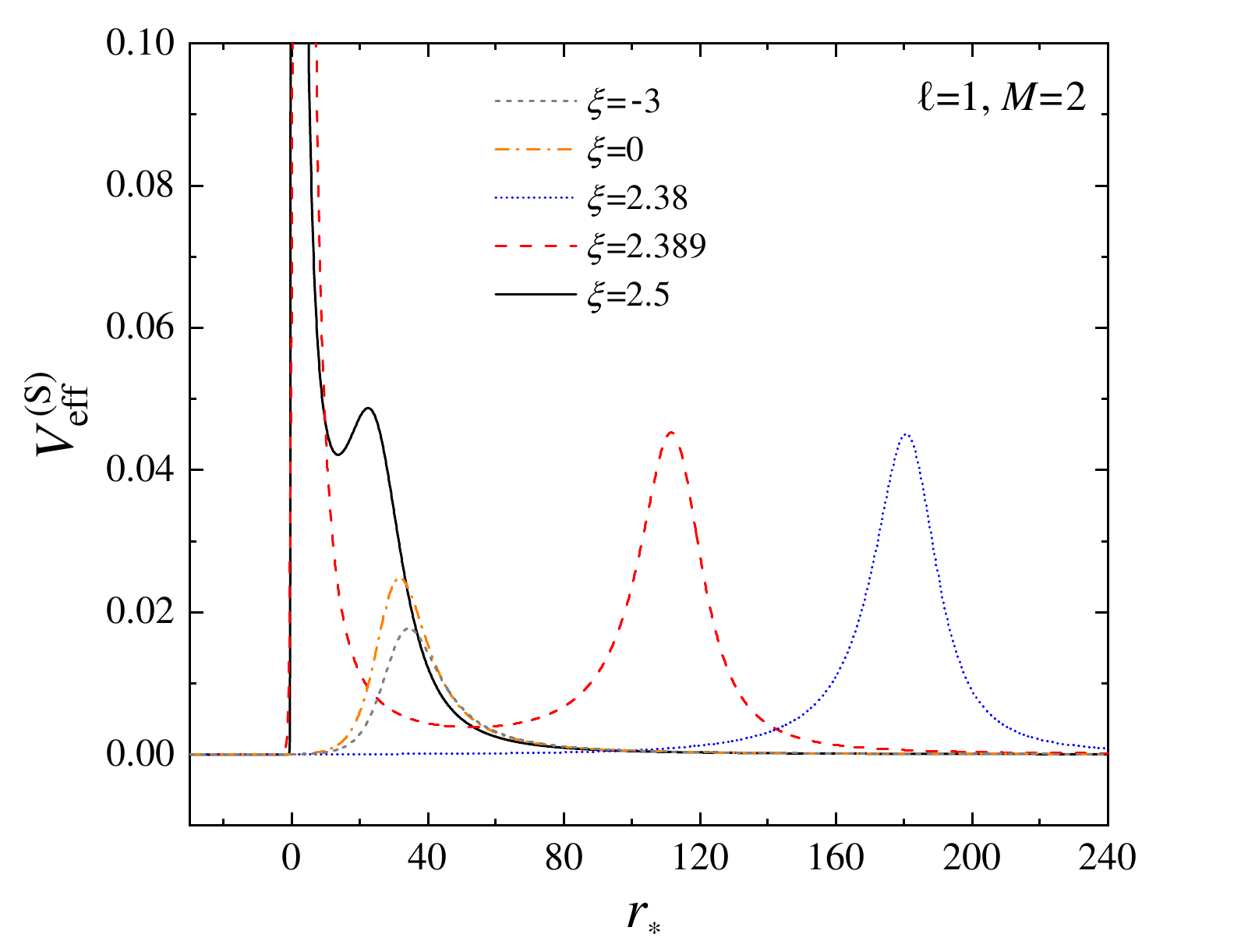}
  \caption{Representative scalar-field effective potentials as functions of the radial coordinate $r_*$ for different values of the black-hole mass $M$, scalar-hair parameter $\xi$, and multipole number $\ell$. The panels illustrate both the standard single-barrier shape and the parameter range where an additional maximum appears outside the horizon.}%
    \label{fig:VeffSF}
\end{figure*}
\section{Results}

We now present the numerical results for wave propagation in the parity-symmetric beyond-Horndeski geometry. We begin with the structure of the effective potentials and the quasinormal spectra, and then turn to the scattering problem, graybody factors, and Hawking emission.

\subsection{QNMs spectra}
The typical form of the effective potentials for the SF case is shown in Fig.~\ref{fig:VeffSF}. For the vector and Dirac fields, the qualitative behavior is similar to that of the SF case with $\ell>0$. The corresponding potentials are positive definite and contain a single peak. Additionally, there exists a domain of parameters $(M,\xi)$ for which the potentials have two maxima, as shown in Fig.~\ref{fig:VeffSF} (right panel). For the $\ell=0$ SF mode, the effective potential may contain a negative gap. However, this does not imply the existence of unstable modes. Using the $S$-deformation method~\cite{Konoplya:2011qq}, with $S=-f(r)/r$, one obtains the deformed potential $V_{\rm eff}=f(r)\ell(\ell+1)/r^2\ge0$, which is non-negative. Therefore, no unstable modes are present.

\begin{figure*} 
\includegraphics[width=0.33\linewidth]{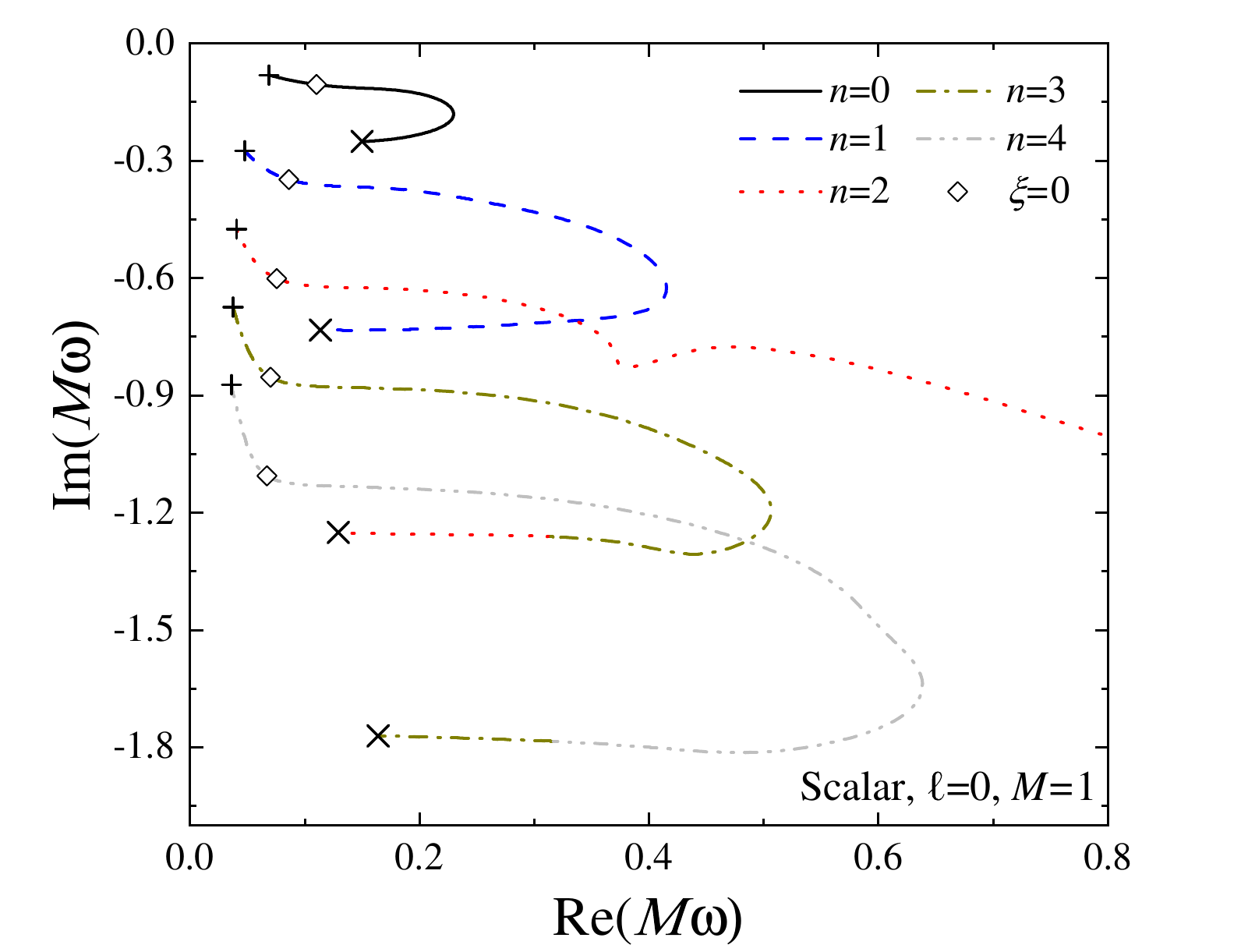}%
\includegraphics[width=0.33\linewidth]{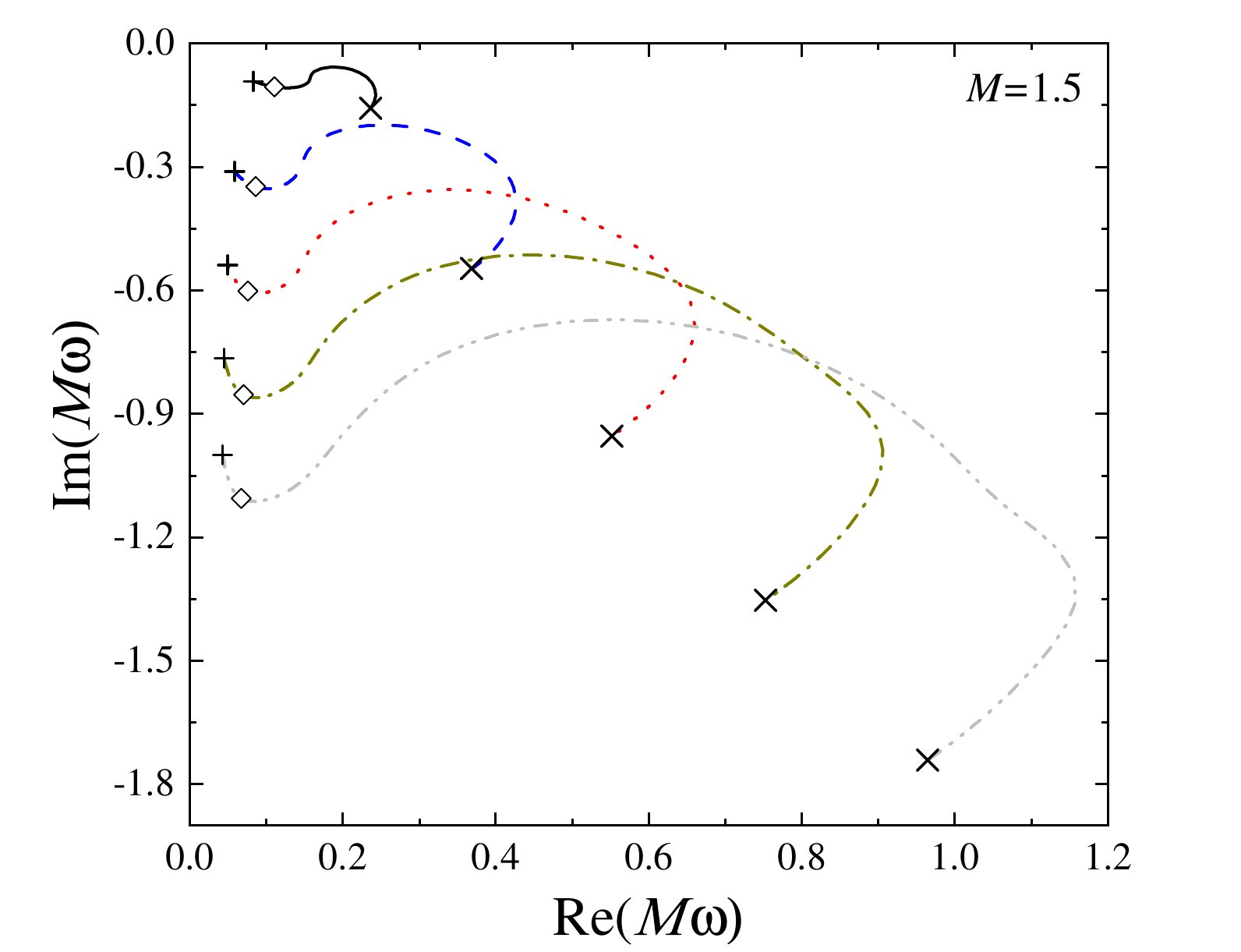}%
\includegraphics[width=0.33\linewidth]{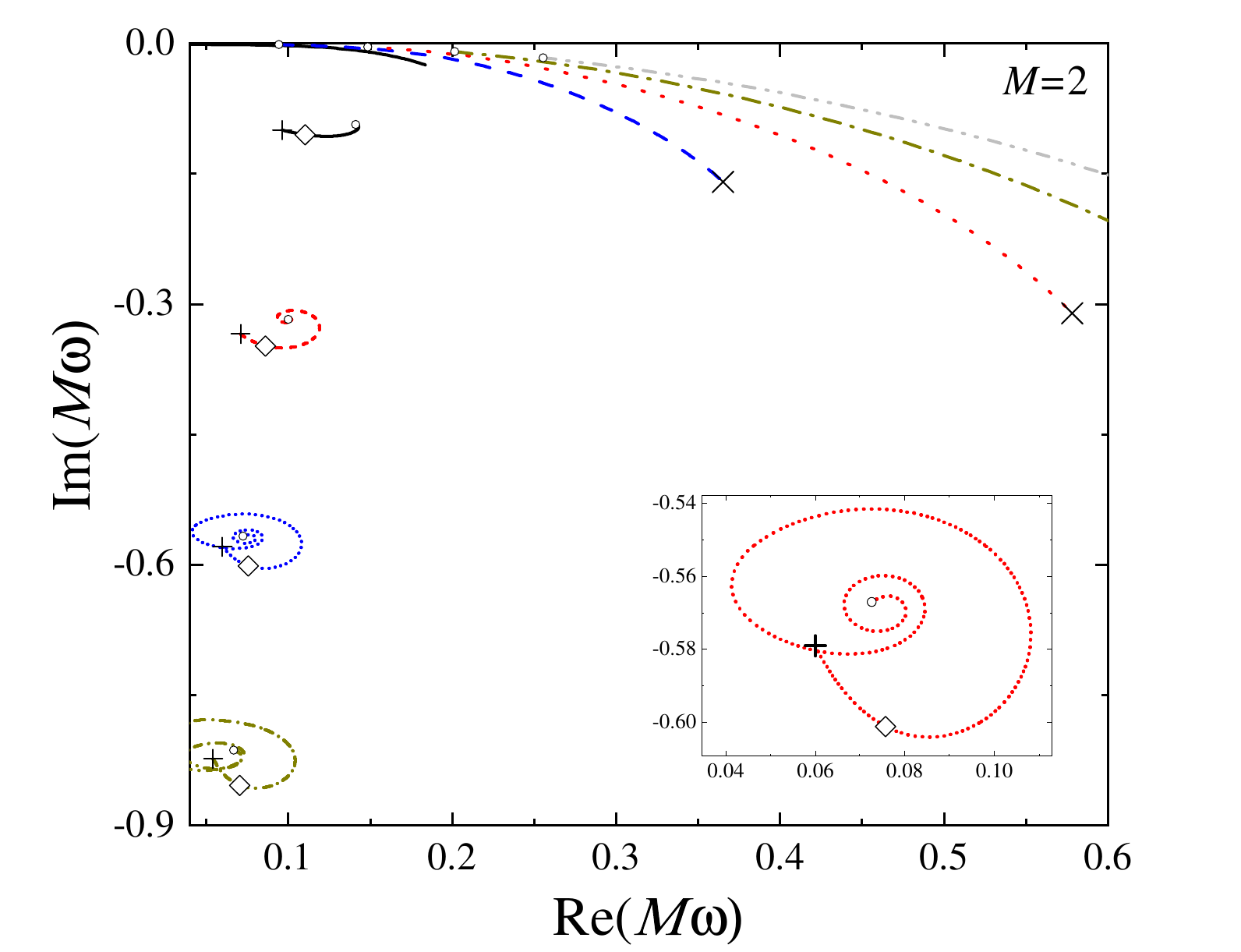}\\

    \caption{Quasinormal-mode branches of the massless scalar test field for representative masses and $\xi\in(-2,\xi_{\rm cr})$, shown for the monopole mode $\ell=0$. The rhombus marks the Schwarzschild limit $\xi=0$, the plus and cross denote the left and right ends of the scanned interval, and the circle indicates the horizon-branch transition where the relevant horizon radius jumps.}%
    \label{fig:qnms_scalar}
\end{figure*}

\begin{figure*}
\includegraphics[width=0.33\linewidth]{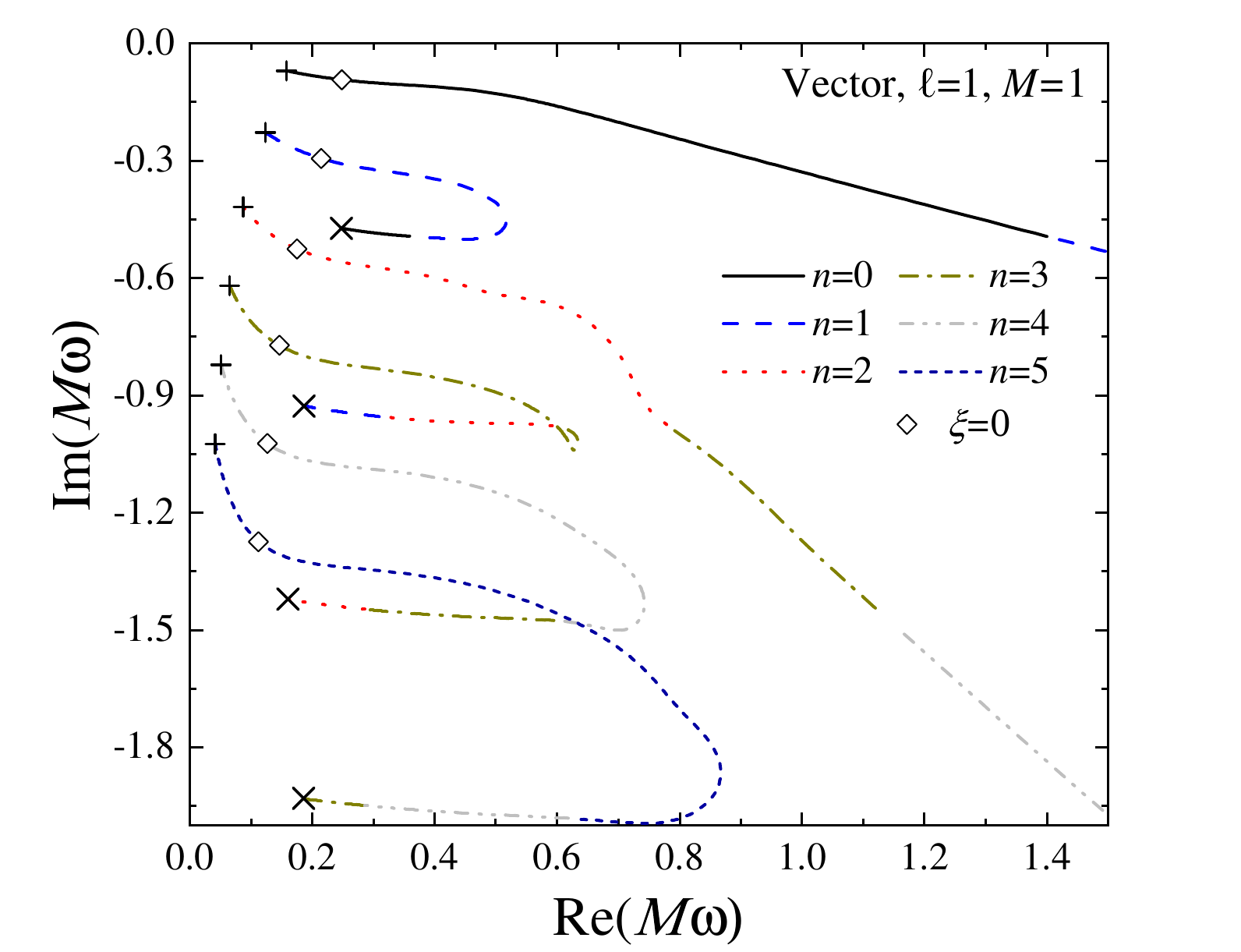}%
\includegraphics[width=0.33\linewidth]{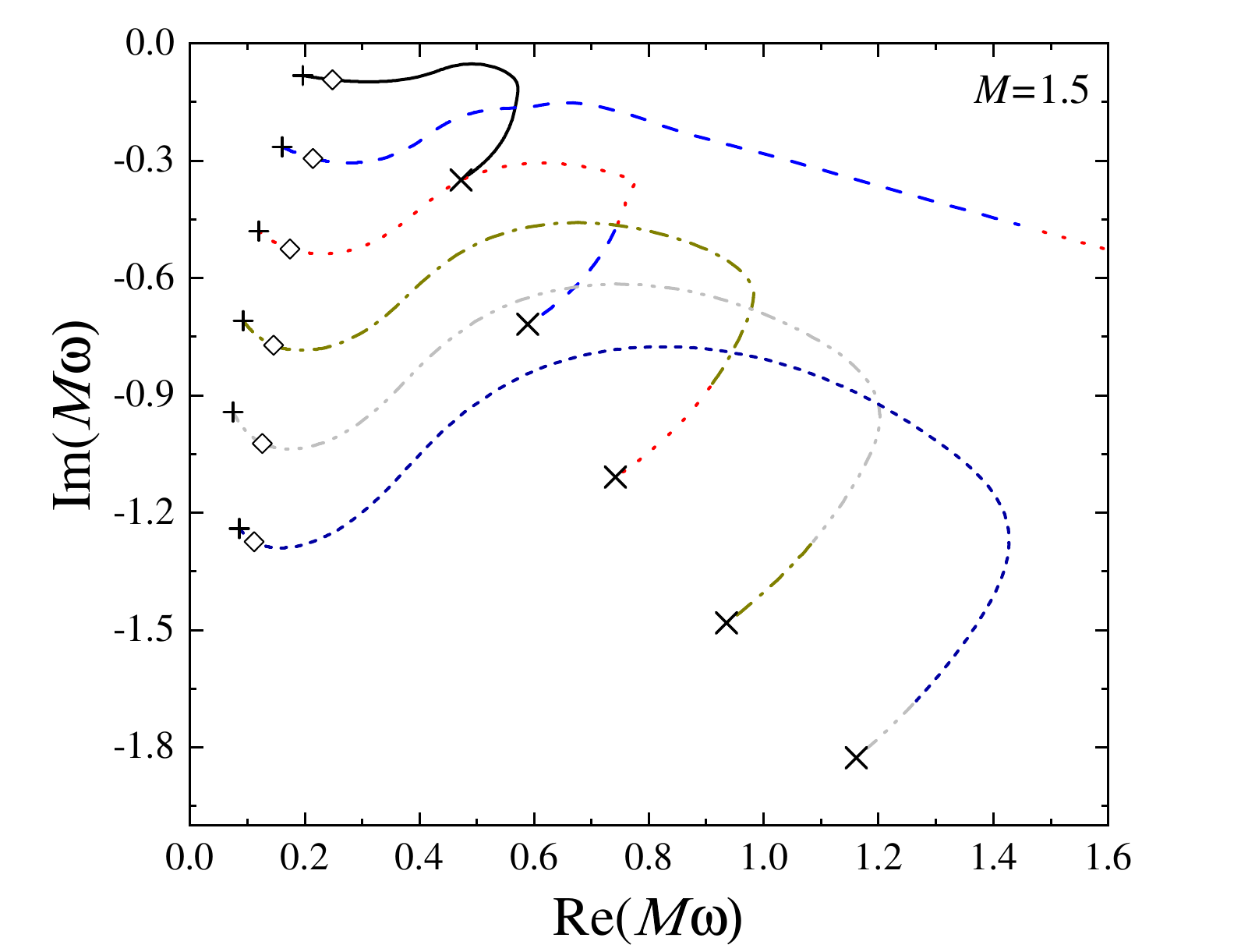}%
\includegraphics[width=0.33\linewidth]{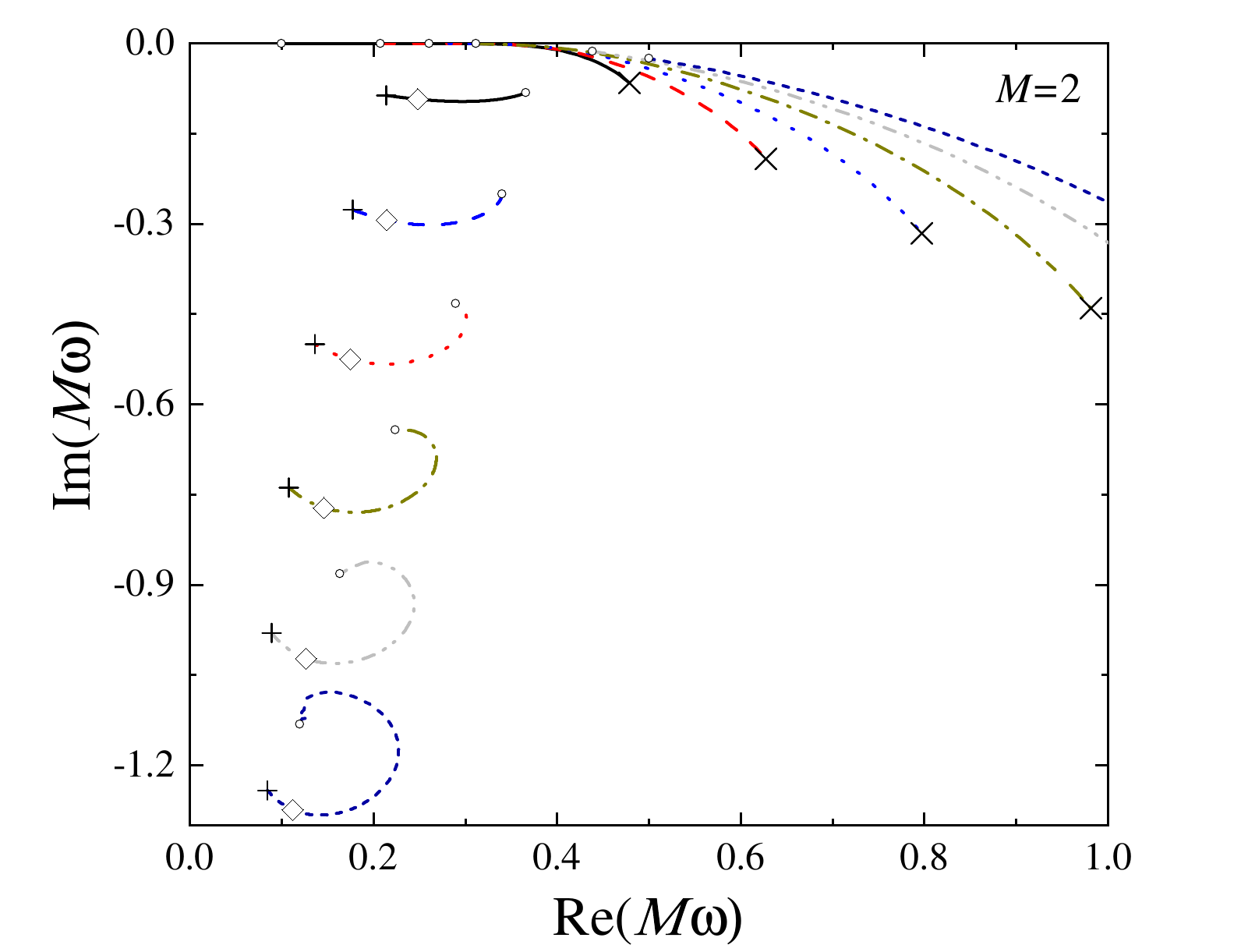}
\caption{Quasinormal-mode branches of the electromagnetic test field for representative masses and $\xi\in(-2,\xi_{\rm cr})$, shown for the lowest allowed vector multipole $\ell=1$. The rhombus marks the Schwarzschild limit $\xi=0$, the plus and cross denote the left and right ends of the scanned interval, and the circle indicates the horizon-branch transition where the relevant horizon radius jumps.}%
    \label{fig:qnms_vector}
\end{figure*}

\begin{figure*}
\includegraphics[width=0.33\linewidth]{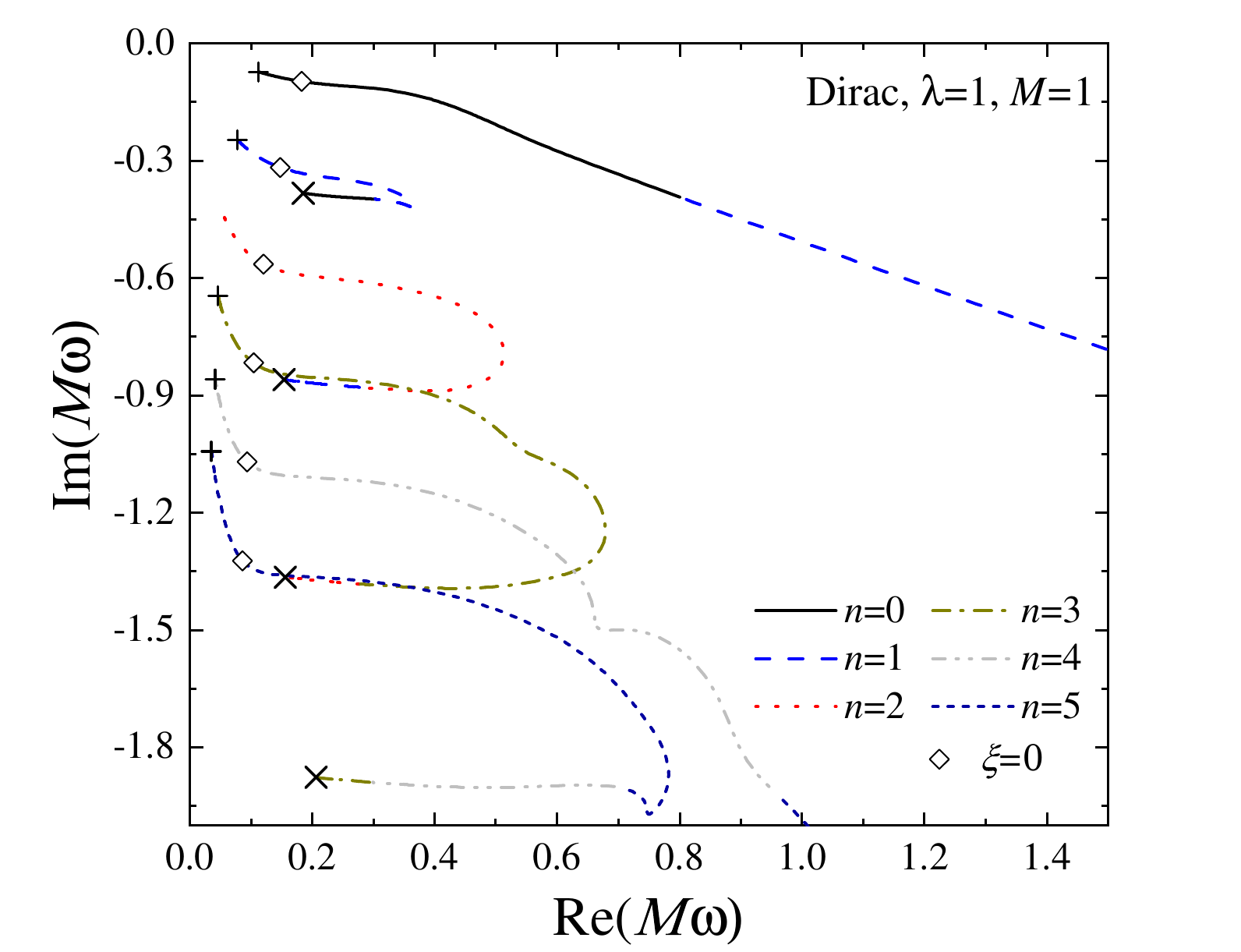}%
\includegraphics[width=0.33\linewidth]{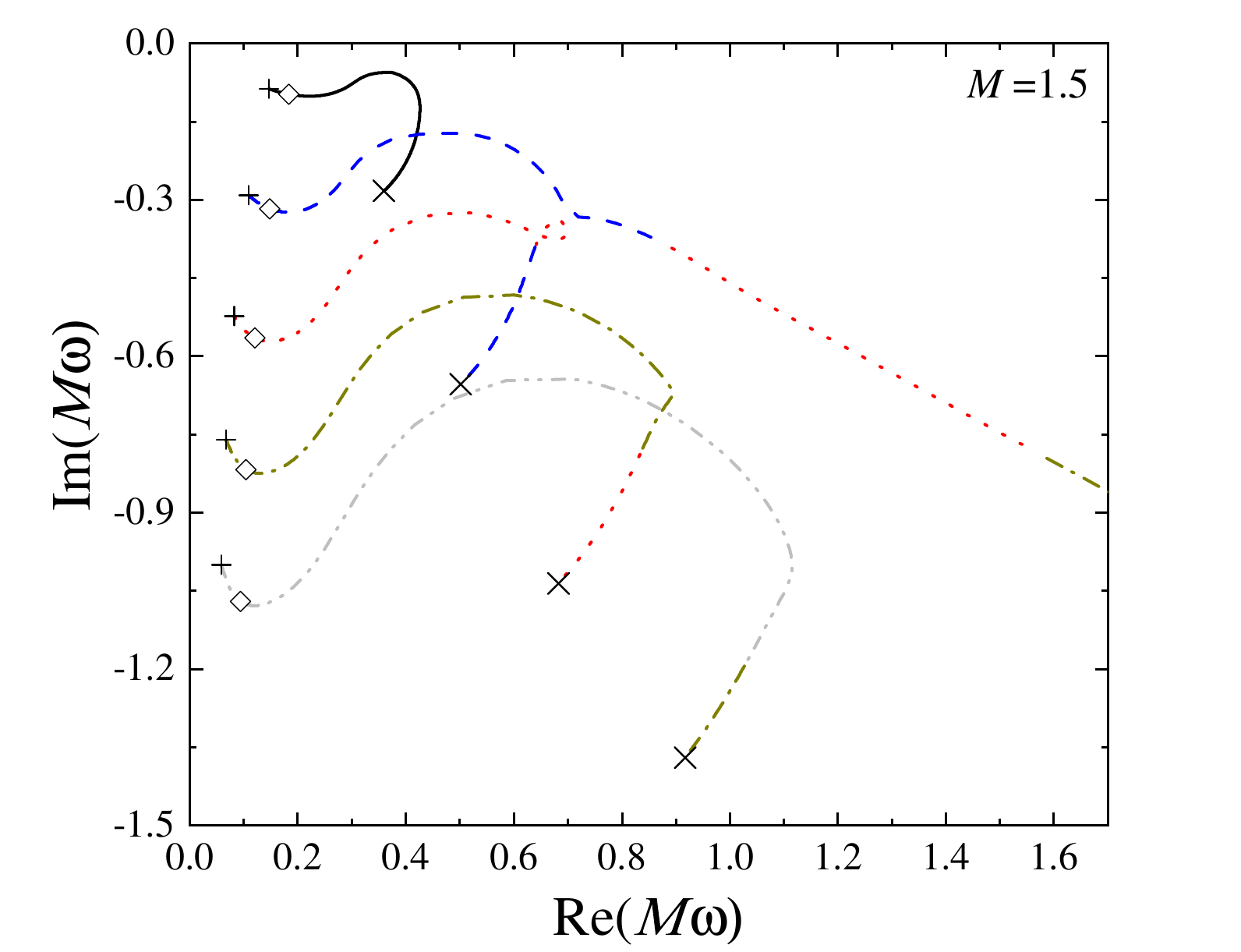}%
\includegraphics[width=0.33\linewidth]{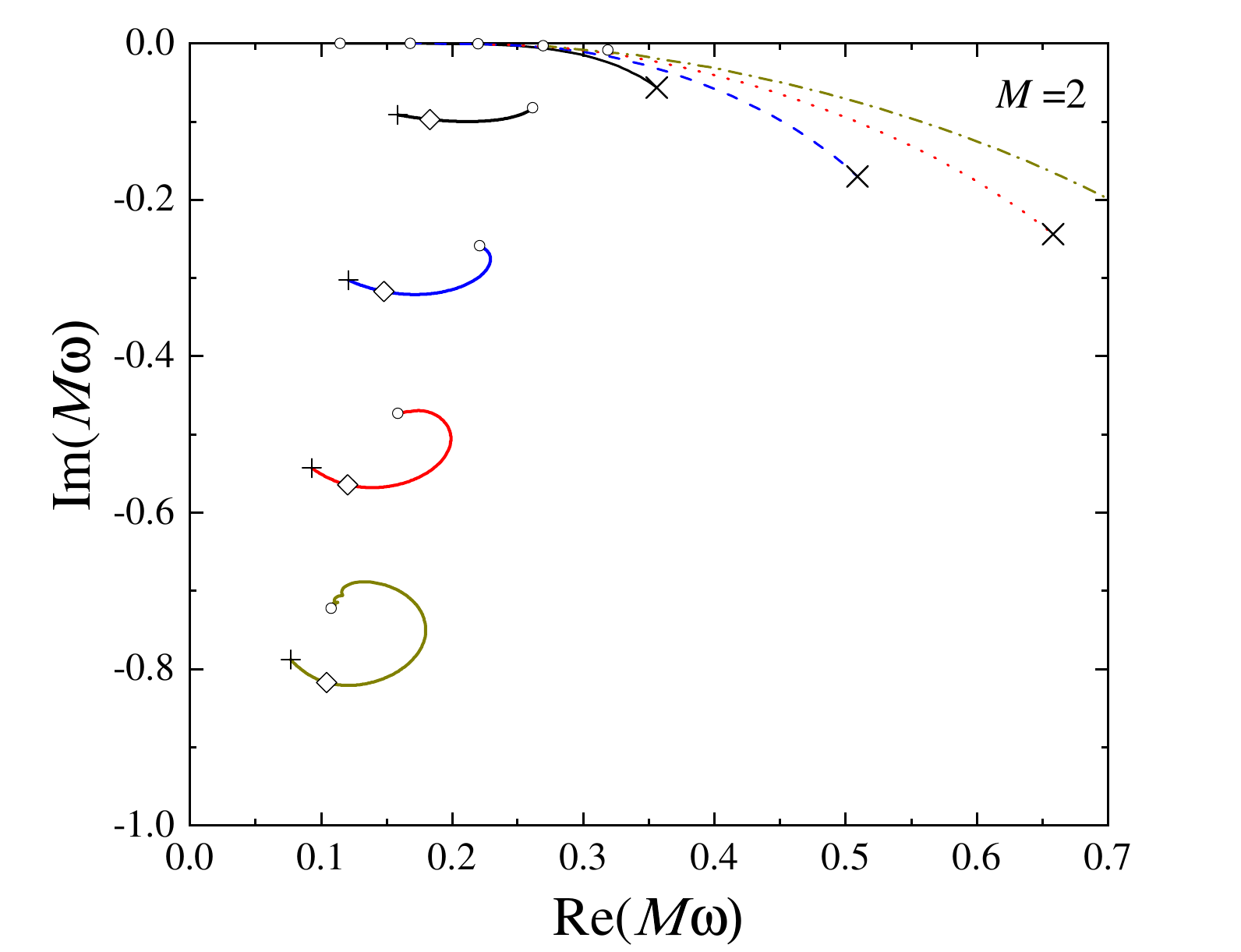}
    \caption{Quasinormal-mode branches of the massless Dirac test field for representative masses and $\xi\in(-2,\xi_{\rm cr})$, shown for the lowest angular mode used in the numerical scan. The rhombus marks the Schwarzschild limit $\xi=0$, the plus and cross denote the left and right ends of the scanned interval, and the circle indicates the horizon-branch transition where the relevant horizon radius jumps.}%
    \label{fig:qnms_dirac}
\end{figure*}

We numerically calculated QNMs spectra for different values of solutions parameters.  The main results are shown in Figs. \ref{fig:qnms_scalar}, \ref{fig:qnms_vector}, and  \ref{fig:qnms_dirac} for lowest $\ell$ for the fundamental mode and first few overtones.
In these figures the Schwarzschild value $\xi=0$ is used as a reference point, while the other markers indicate the endpoints of the scanned interval and, when present, the transition point at which the relevant horizon branch changes. The curves should therefore be read as continuous spectral branches followed in $\xi$, rather than only as an ordering by the magnitude of ${\rm Im}\,\omega$.

For $\xi<0$, the QNM frequencies tend to zero in the limit $\xi\to-\infty$. This behavior can be seen analytically in the eikonal regime $\ell\to\infty$, where QNM frequencies are related to the parameters of the unstable circular null orbit through the eikonal correspondence~\cite{Cardoso:2008bp}.

Let $\alpha=-\xi>0$. The metric function can be written as $f(r)=1-2M/r-\alpha h(r)$, where $h(r)=\arctan(1/r)/r+1/(1+r^2)$. The outer horizon $r_h(\alpha)$ is a strictly monotonic function of $\alpha$. Indeed, from $f(r_h)=0$ one has $\alpha=g(r)/h(r)\equiv A(r)$, where $g(r)=1-2M/r$. For $r>2M$, $g(r)>0$, $g'(r)>0$, $h(r)>0$, and $h'(r)<0$. Hence $A'(r)>0$. Since $A(2M)=0$ and $A(r)\to\infty$ as $r\to\infty$, $A$ maps $(2M,\infty)$ monotonically onto $(0,\infty)$. Therefore $r_h(\alpha)=A^{-1}(\alpha)$ is also strictly increasing. For large $\alpha$, one obtains
\eq{
 r_h= M+\sqrt{2\alpha}+\frac{3M^2-2}{6\sqrt{2\alpha}}+O\left(\frac{1}{\alpha}\right),~ \alpha\to\infty,}
or simply $r_h\sim\sqrt{2\alpha}$.

The photon-sphere radius $r_{\rm ph}$ is determined by  root of \cite{Claudel_2001}: $rf'(r)-2f(r)=0$. Since $r_{\rm ph}>r_h\sim\sqrt{2\alpha}$, one has $r_{\rm ph}\to\infty$ as $\alpha\to\infty$. Using the large-$r$ asymptotics of $f(r)$, we obtain
\eq{
 r_{\rm ph}=2\sqrt{\alpha}+\frac{3M}{2}+\frac{9M^2-4}{16\sqrt{\alpha}}+O\left(\frac{1}{\alpha}\right),~ \alpha\to\infty .
}
Thus $r_{\rm ph}\sim2\sqrt{\alpha}$ and $r_{\rm ph}/r_h\to\sqrt{2}$.

This yields that  eikonal QNMs of test fields are 
\eq{
 \omega\sim
 \frac{\ell+\frac12}{2\sqrt{2\alpha}}
 -i\,\frac{n+\frac12}{2\sqrt{\alpha}},
 \qquad \alpha\to\infty,
}
Hence $\omega$ values are pushed toward the origin in the complex-frequency plane. We need to notice  that such a correspondence may not be valid for gravitational sector, as shown in \cite{Konoplya:2017wot,Konoplya:2025afm}.

\begin{figure*}
\includegraphics[width=0.33\linewidth]{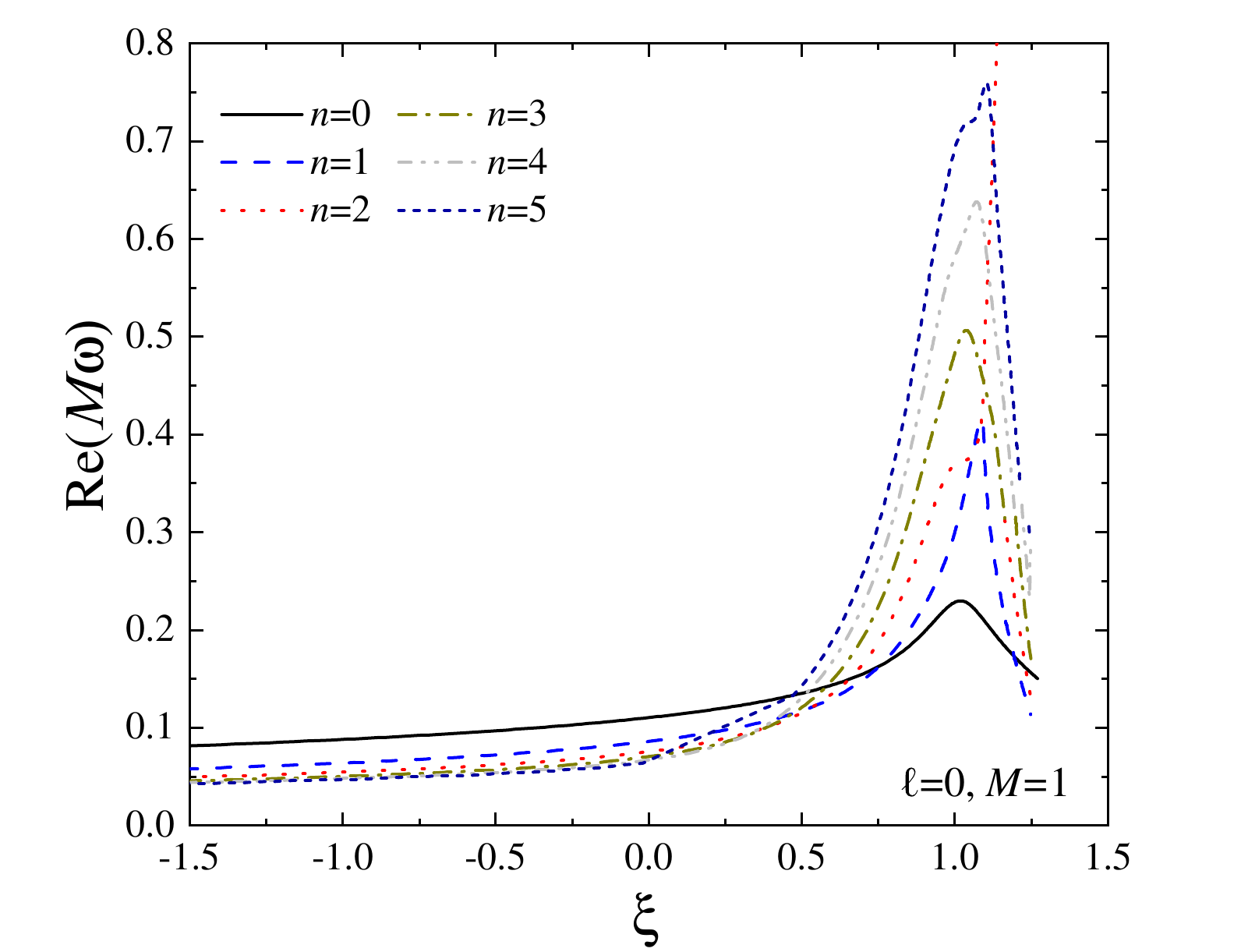}%
\includegraphics[width=0.33\linewidth]{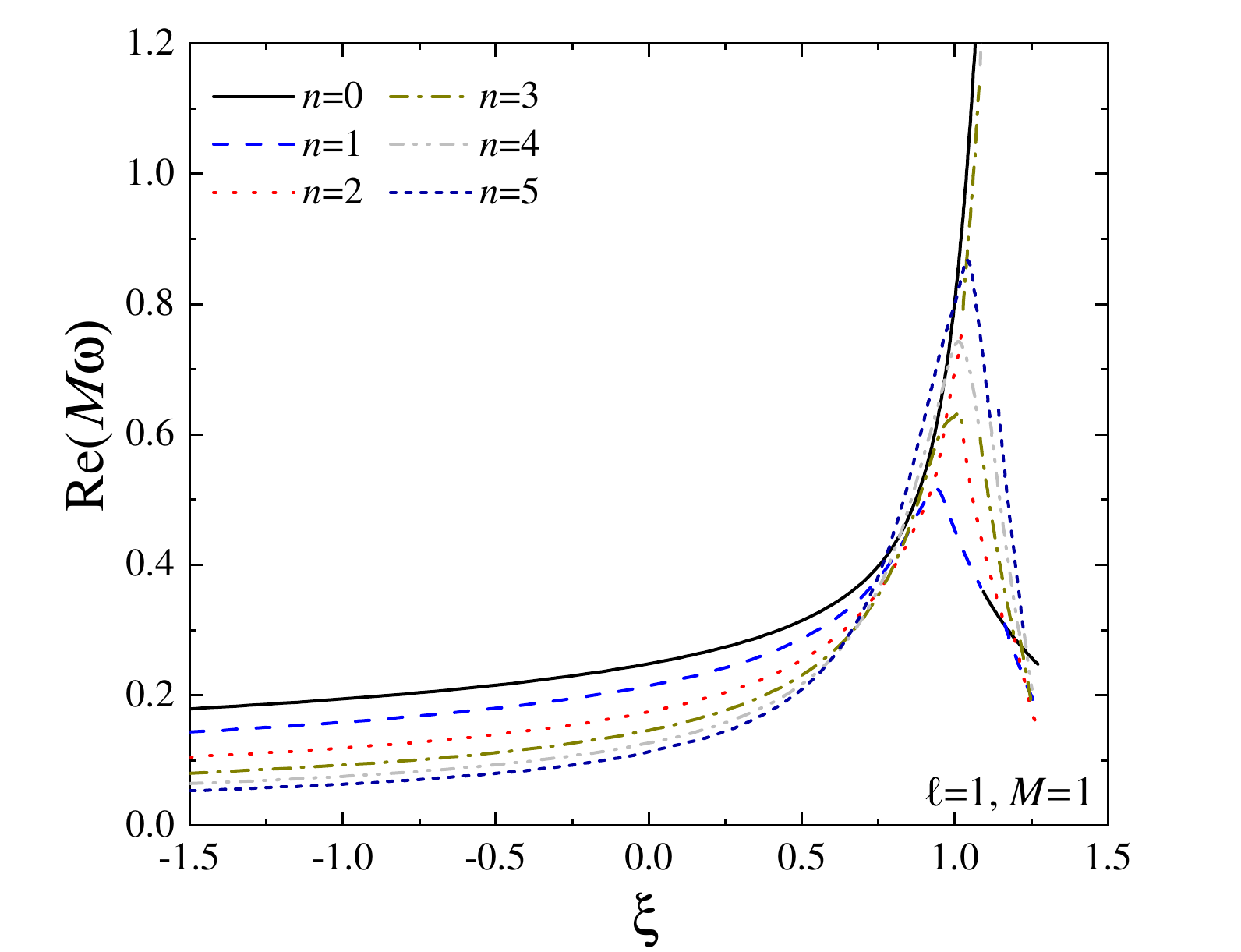}%
\includegraphics[width=0.33\linewidth]{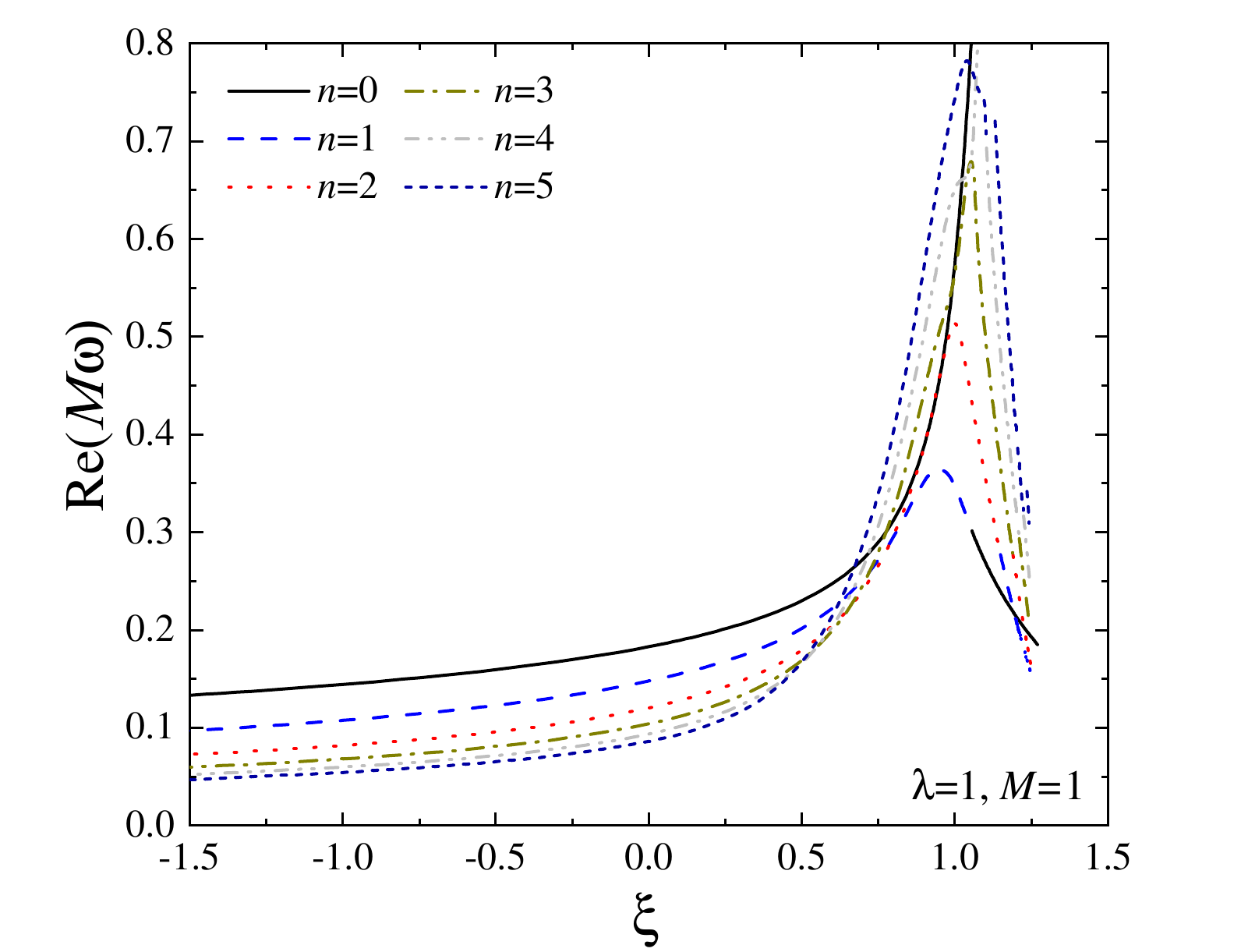}
    \caption{Real parts of representative quasinormal frequencies for scalar, electromagnetic, and Dirac test fields at $M=1$ as functions of $\xi\in(-1.5,\xi_{\rm cr})$.}%
    \label{fig:appendix_qnms_re}
\end{figure*}

\begin{figure*}
\includegraphics[width=0.33\linewidth]{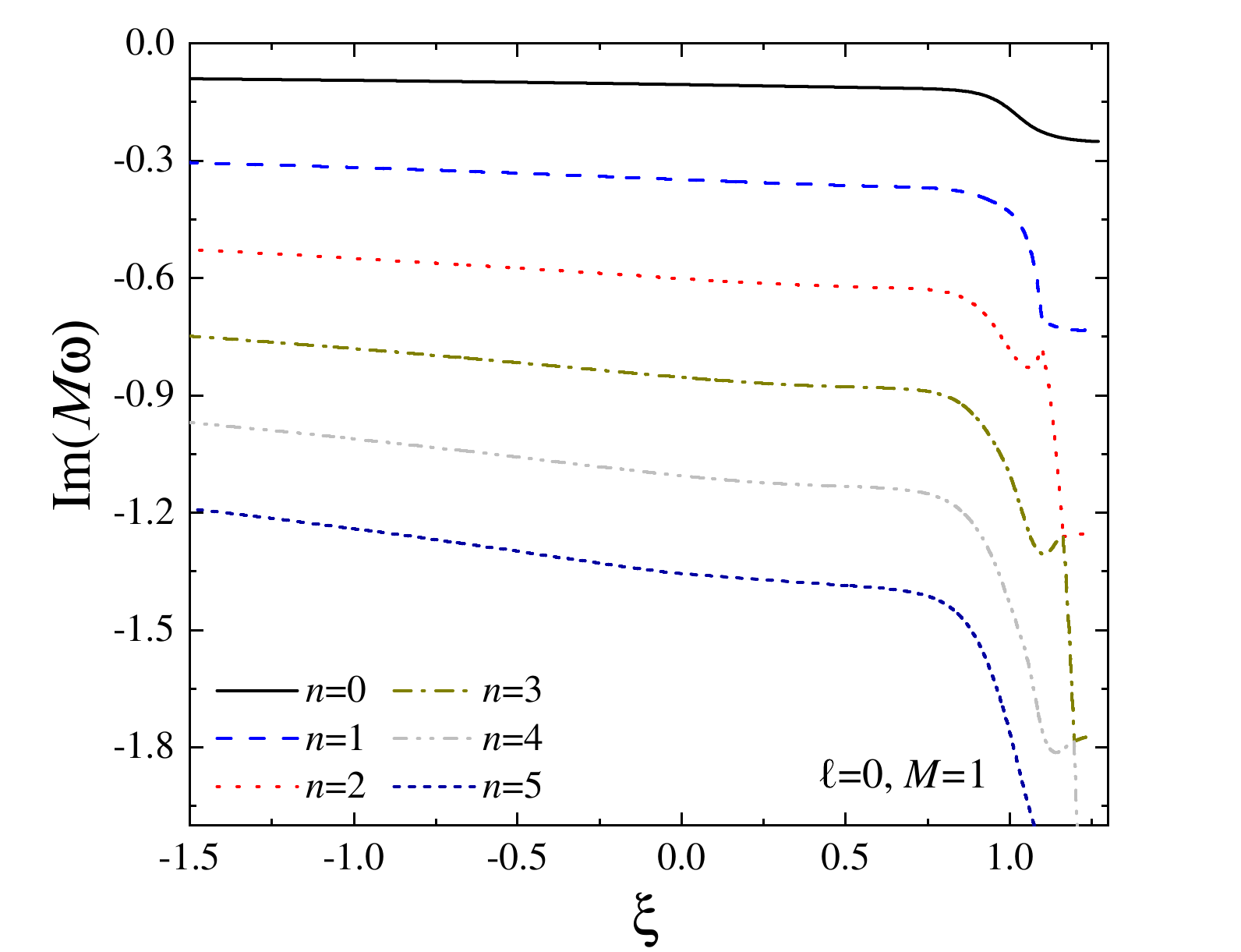}%
\includegraphics[width=0.33\linewidth]{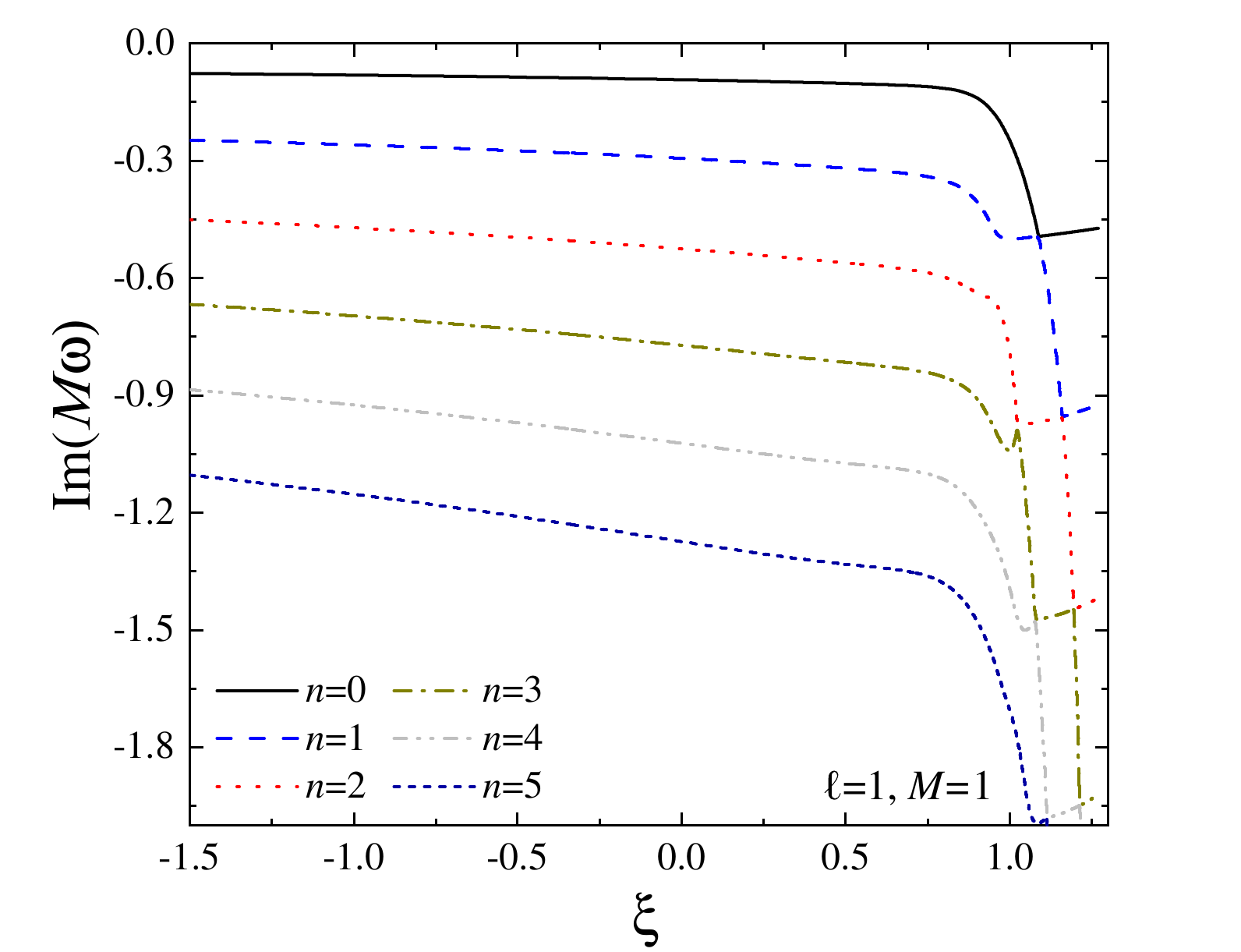}%
\includegraphics[width=0.33\linewidth]{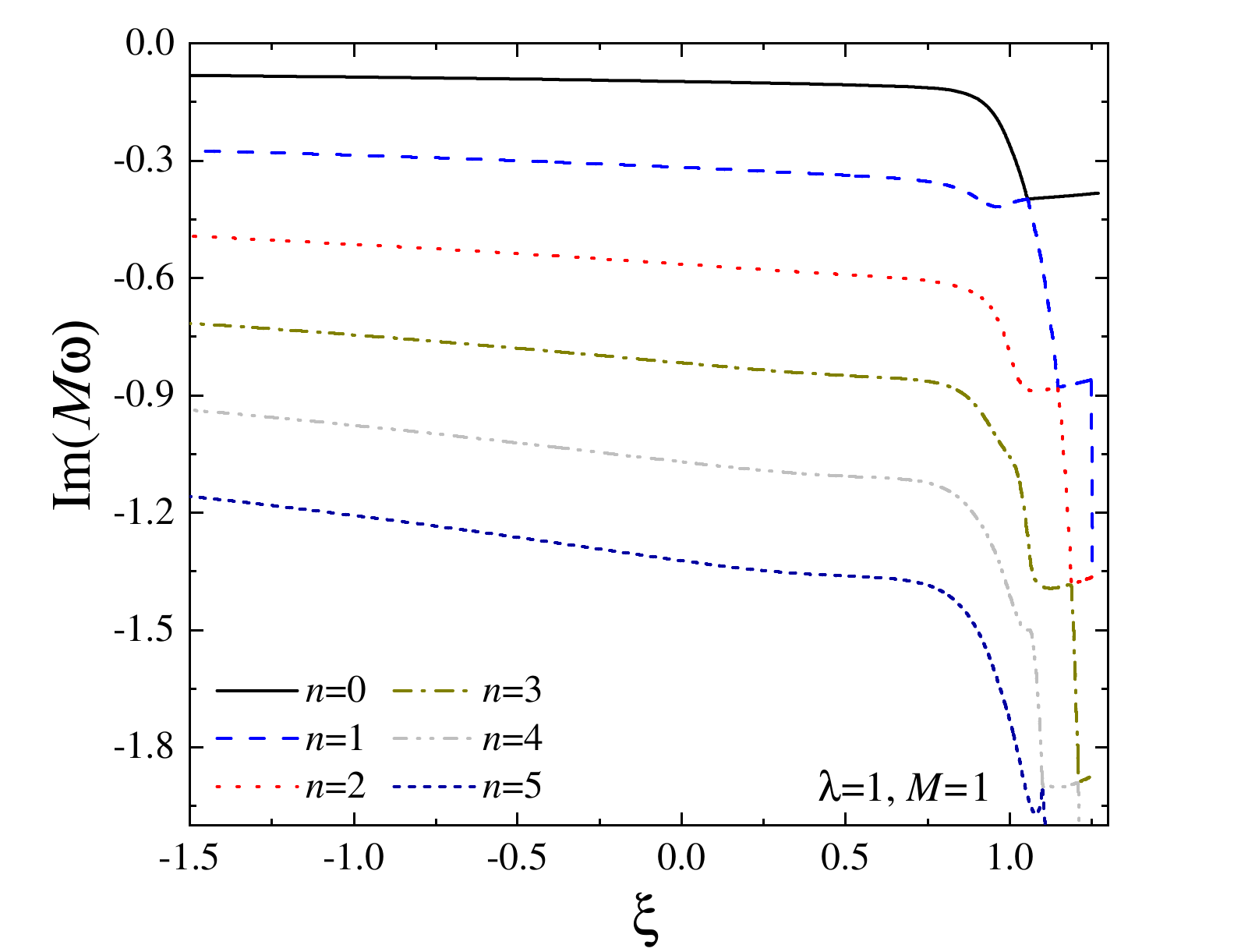}
    \caption{Imaginary parts of representative quasinormal frequencies for scalar, electromagnetic, and Dirac test fields at $M=1$ as functions of $\xi\in(-1.5,\xi_{\rm cr})$. }%
    \label{fig:appendix_qnms_im}
\end{figure*}

For $\xi>0$, such simple dependencies are absent. In this case, there exists a set of specific values of $\xi$, depending on $M$, that separate qualitatively different behaviours, as well as a critical value $\xi=\xi_{\rm cr}(M_{\rm cr})$, that separates BHs from ECOs. For example, for the scalar field case with $M=1$, Figs.~\ref{fig:qnms_scalar}--\ref{fig:qnms_dirac}, and Fig. \ref{fig:appendix_qnms_re}, \ref{fig:appendix_qnms_im}  show that the modes initially move slowly away from their Schwarzschild counterparts up to some value $\xi<\xi_1$. In the range $\xi_1<\xi<\xi_{\rm cr}$, the deviations become more prominent, and a sequence of mode switchings appears (see Fig. \ref{fig:appendix_qnms_im}). In this interval, the effective potential gradually deforms, additional extrema tend to appear, and neighboring QNM branches may approach each other and exchange their ordering by damping rate. In the vector field case with $M=1$, the fundamental mode switches with the first overtone, and analogous switchings occur for higher overtones, whereas for the third and higher overtones one initially observes eigenvalue repulsion followed by switching with other modes (see Fig. \ref{fig:appendix_qnms_im}, middle panel). For the scalar field, there is no fundamental-mode switching, but switching appears for higher overtones (see Fig. \ref{fig:appendix_qnms_im}, left panel). The same qualitative behavior is also present for the other fields. Increasing $M$ may remove the mode switching for the first overtones and shift it to larger overtone numbers. Additionally, as $\xi\to\xi_{\rm cr}$, then $|{\rm Im}\,\omega|$ starts to increase rapidly.

\begin{figure*}
\includegraphics[width=0.49\linewidth]{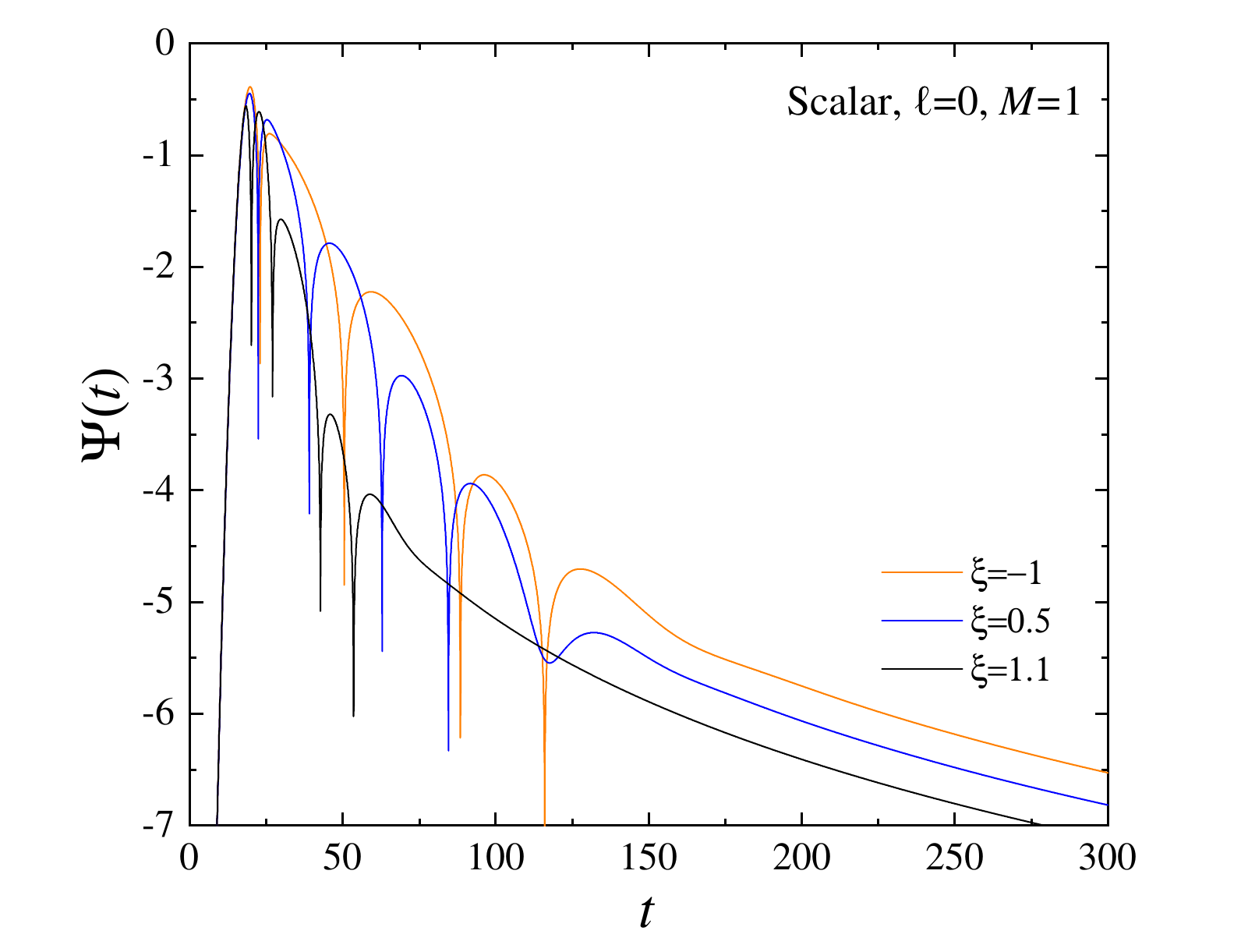}%
\includegraphics[width=0.49\linewidth]{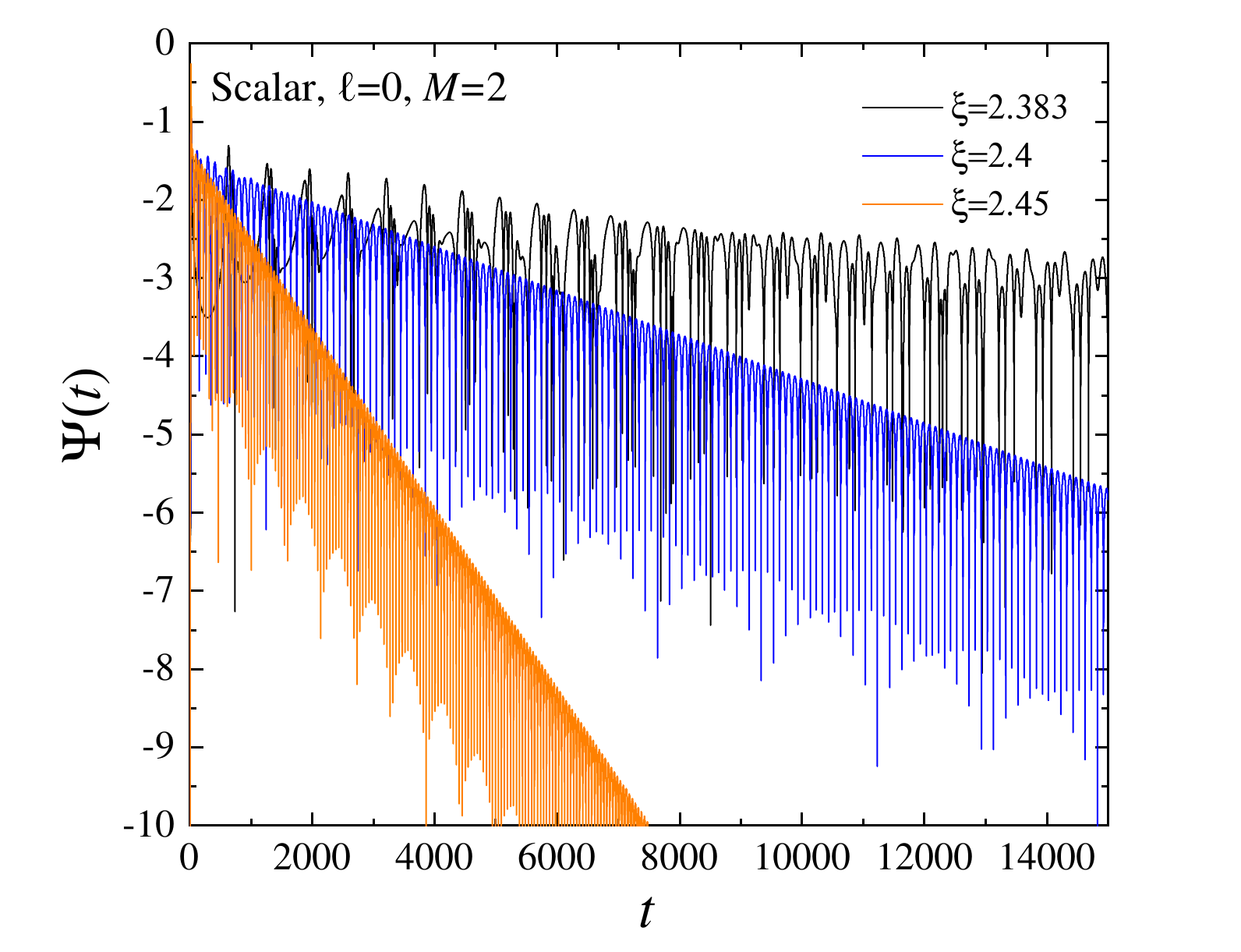}%
    \caption{Representative scalar-field time-domain waveforms for two choices of the mass and for different  values of $\xi$. The profiles illustrate the transition from ordinary ringdown to signals with delayed echo-like contributions when a double-barrier structure is present.}%
    \label{fig:waveSol}
\end{figure*}

For $M>M_1$, at some $\xi=\xi_t$, the inner black-hole horizon, which is the first one among the three horizons, becomes the single horizon, i.e. $r_h(\xi_t-0)>r_h(\xi_t+0)$. This immediately leads to the formation of an additional barrier that was initially hidden under the horizon, and to the appearance of trapped modes and echoes. Examples of time-domain waveform profiles are shown in Fig.~\ref{fig:waveSol}. The spectrum becomes discontinuous because the height of the first barrier is significantly larger than that of the second one, so it effectively serves as a potential wall. This leads to a spectrum similar to the ECO case~\cite{Cardoso:2016rao,Cardoso:2019rvt}, where the dominant frequencies correspond to long-lived modes. In the SF case with $\ell=0$, one also observes a peculiar spiral-like QNM curve caused by the variation of the distance between the first and second barriers. Similar behaviour appears for higher overtones for the other field cases.

In this regime the exterior problem is better viewed as a finite cavity bounded by two potential barriers. The real part of the trapped frequencies is approximately set by the cavity size in tortoise coordinate, while the imaginary part is controlled by leakage through the outer barrier. As the separation between the barriers changes with $\xi$, the phase accumulated in the cavity changes as well, producing the spiral-like trajectories and the echo pattern in the time-domain response. Closely related double-barrier spectra and echoes occur for exotic compact objects and for hairy black-hole effective potentials~\cite{Cardoso:2016rao,Cardoso:2019rvt,Guo:2022sfl, Konoplya:2025uiq, Konoplya:2025hgp}. 

For completeness, we also demonstrate the time-domain profiles in Fig.~\ref{fig:waveSol}  to illustrate the waveform response.

For fixed  $\xi\neq 4M/\pi$ and $M>M_{\rm cr}$, the space-time contains a point-like naked singularity at the center. Since the space-time is not globally hyperbolic, Cauchy data on a regular space-like hypersurface do not determine a unique evolution. Following \cite{Wald}, this problem can be resolved if the spatial part $A$ of the evolution operator (\ref{eq:generic-master}), ($\partial_t^2\Psi=-A\Psi$), admits a unique self-adjoint extension $A_E$. In this case the boundary condition at the singularity is fixed uniquely, and the dynamics is well-defined. If the extension $A_E$ is not unique, different self-adjoint extensions correspond to different boundary conditions at the singularity and lead to inequivalent dynamics.

To study essential self-adjointness of $A$, we need compute the deficiency indices of $A$ in $L^2((0,\infty),dr_*)$.\footnote{For a symmetric operator $A$ in $L^2((0,\infty),dr_*)$, the deficiency indices are $n_\pm=\dim\ker(A^*\mp i)$, where $A^*$ is the adjoint operator. Essential self-adjointness is equivalent to $n_+=n_-=0$.}
Using the asymptotic expansion of $f(r)$ near the singularity, one obtains the following asymptotic behaviour of the effective potentials
\eq{
V^{(S)}_{\rm eff}\sim -\frac{\delta^2}{4 r^4}, ~V^{(V)}_{\rm eff}\sim \frac{\ell (\ell+1) \delta}{2 r^3},~V^{(D)}_{\rm eff}\sim-\frac{3 \delta ^{3/2} \lambda }{4 \sqrt{2}r^{7/2}},
}
where $\delta=\pi  \xi-4M>0$\footnote{$\delta$ cannot be negative, because in this case the space-time always has at least one horizon.} and $r^2\simeq\delta \,r_*$, as  $r\to0$. 

Hence
\eq{
V^{(S)}_{\rm eff}\sim -\frac{1}{4 r_*^2}, 
~
V^{(V)}_{\rm eff}\sim \frac{\ell(\ell+1)}{2\sqrt{\delta} r_*^{3/2}},
~
V^{(D)}_{\rm eff}\sim-\frac{3 \lambda }{4 \sqrt{2}
   \sqrt[4]{\delta} r_*^{7/4}}.
}
At spatial infinity, the asymptotic behaviour is the same for all fields,
\eq{V^{(i)}_{\rm eff}=\frac{\ell(\ell+1)}{r^2}+O\left(r^{-3}\right).}
For deficiency index equations $(A^*\pm i)\Psi=0$, the asymptotic solutions near the singularity are
\eq{
\Psi_S \sim C_1 \sqrt{r_*}+C_2 \sqrt{r_*}\ln r_* .
}
for the scalar field, and
\eq{
\Psi_{V,D} \sim C_1+C_2r_* ,}
for the vector and Dirac fields, respectively.
One can see that in all cases $\Psi_i\in L^2((0,\epsilon),dr_*)$ for sufficiently small $\epsilon>0$, $i=(S,V,D)$, i.e. they are square integrable near the singularity.

At spatial infinity, the  asymptotic solution of deficiency index equations can be written as
\eq{
\Psi_i\sim C_+ e^{\kappa_\pm r_*}+C_- e^{-\kappa_\pm r_*}, ~ \kappa_\pm^2=\mp i ,
}
For each choice of sign in the deficiency equation, one of these solutions grows exponentially, whereas the other decays exponentially as $r_*\to\infty$. Hence only one independent solution satisfies $\Psi_i\in L^2((\epsilon,\infty),dr_*)$ for sufficiently large $\epsilon>0$, $i=(S,V,D)$.

Thus the deficiency indices of $A$ are $n_+=n_-=1$ for each type of field. Therefore none of these operators is essentially self-adjoint. Each admits a one-parameter family of self-adjoint extensions, corresponding to different possible boundary conditions at the singularity and different dynamics, respectively. Based on classification proposed in \cite{Horowitz} such naked singularities are quantum mechanically singular for all fields.

In the special case $\xi=4M/\pi$, the solutions describe regular black holes or solitons. For solitons, the center $r=0$ is regular and the dynamics is well-defined. Near the center, one has $V^{(i)}_{\rm eff}\sim \ell(\ell+1)/r_*^2$ and
$
\Psi_i\simeq C_1 r_*^{\ell+1}+C_2 r_*^{-\ell}
$
as $r_*\to0$ for $\ell\ge1$, while for $\ell=0$ one has $\Psi_i\simeq C_1 r_*+C_2$. Regularity at the center requires $C_2=0$, so the admissible solution behaves as $\Psi_i\sim r_*^{\ell+1}$. The analysis of quasinormal spectra for solitons is beyond the scope of this paper.

\subsection{Scattering results}
\begin{figure*}
    \includegraphics[width=0.33\linewidth]{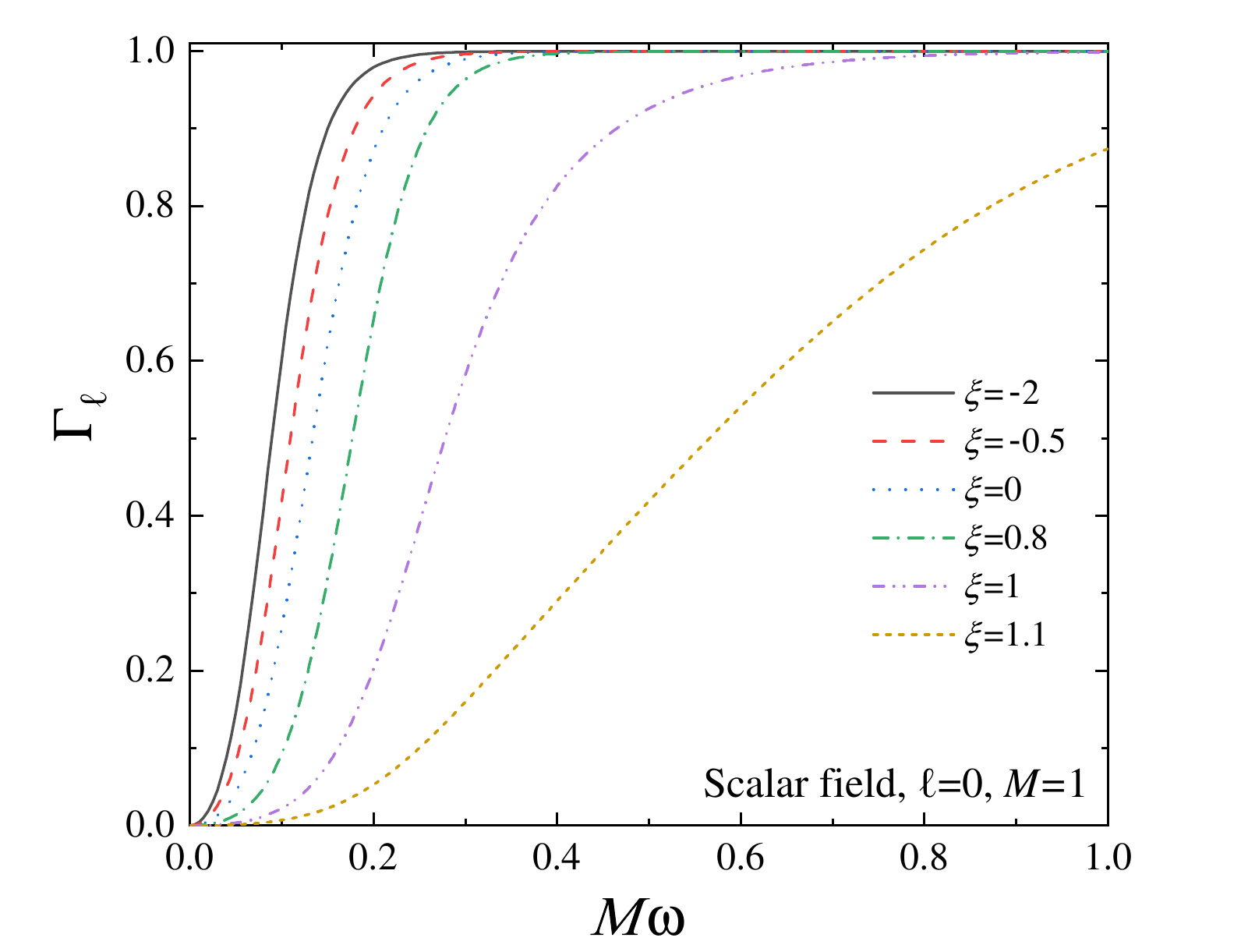}%
    \includegraphics[width=0.33\linewidth]{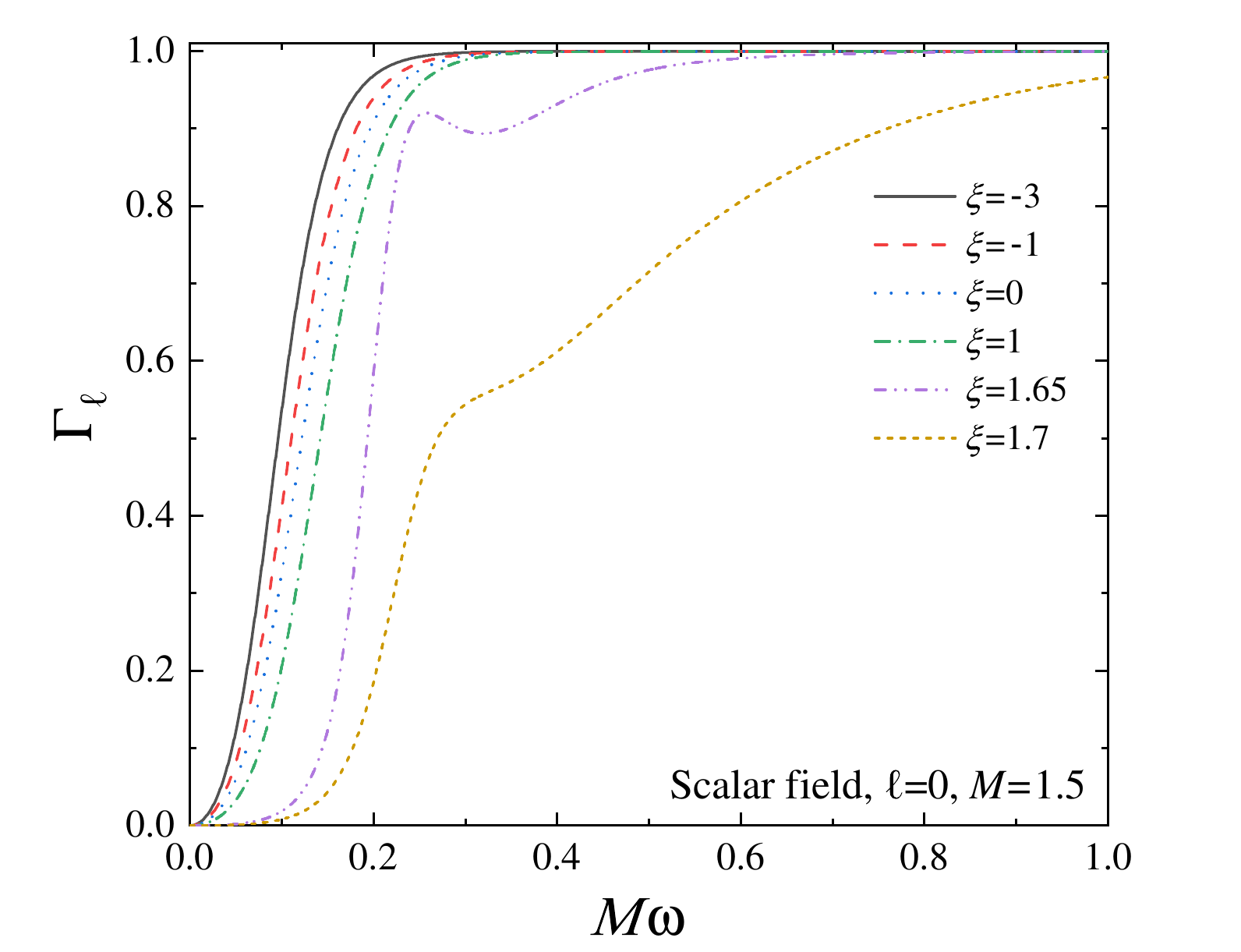}
    \includegraphics[width=0.33\linewidth]{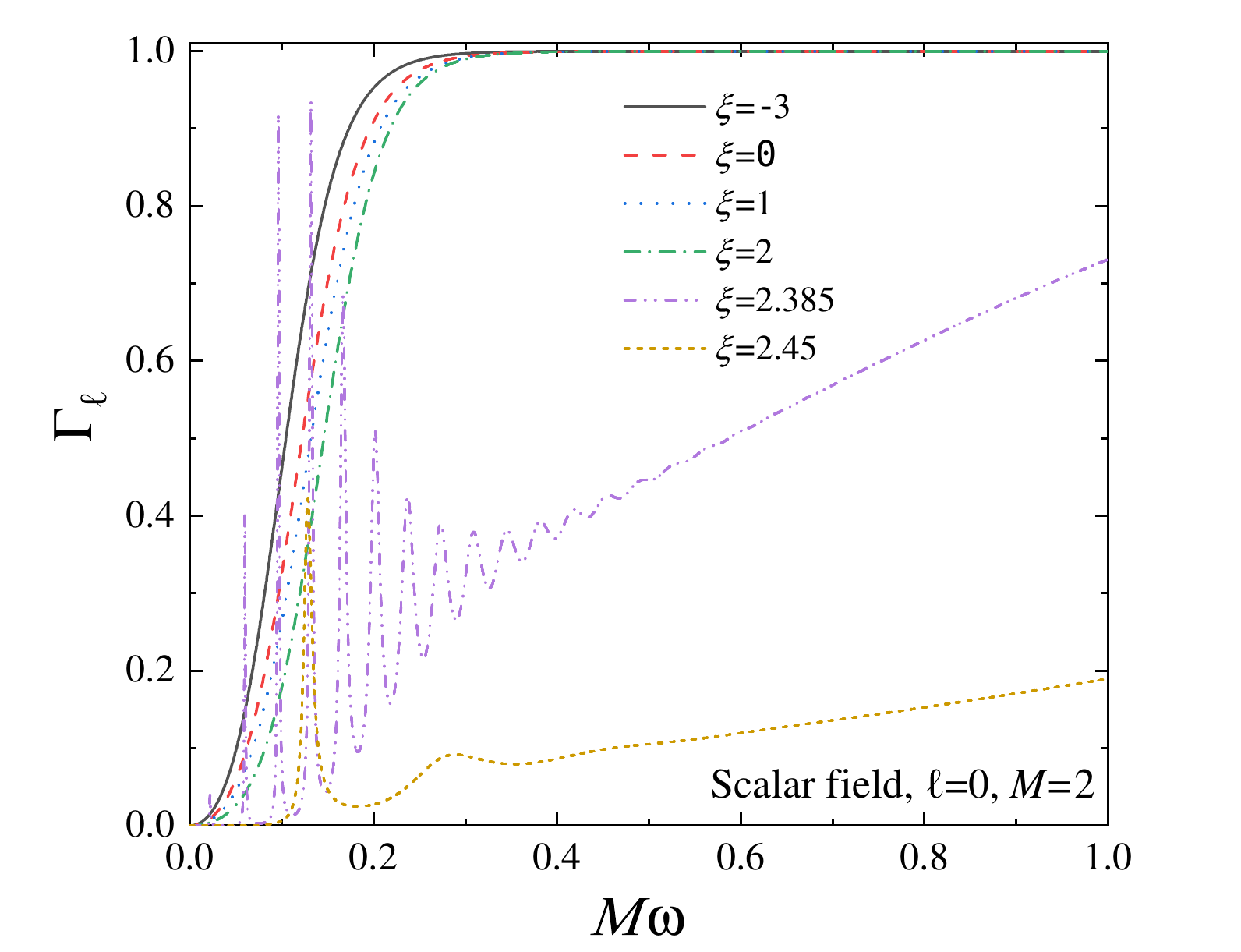}\\

    \caption{Scalar-field graybody factors for the monopole mode $\ell=0$ and representative values of the mass $M$ and hair parameter $\xi$. The curves show how the transmission threshold shifts as the horizon scale and the height of the effective potential change.}%
    \label{fig:GB_scalar}
\end{figure*}

\begin{figure*}
        \includegraphics[width=0.33\linewidth]{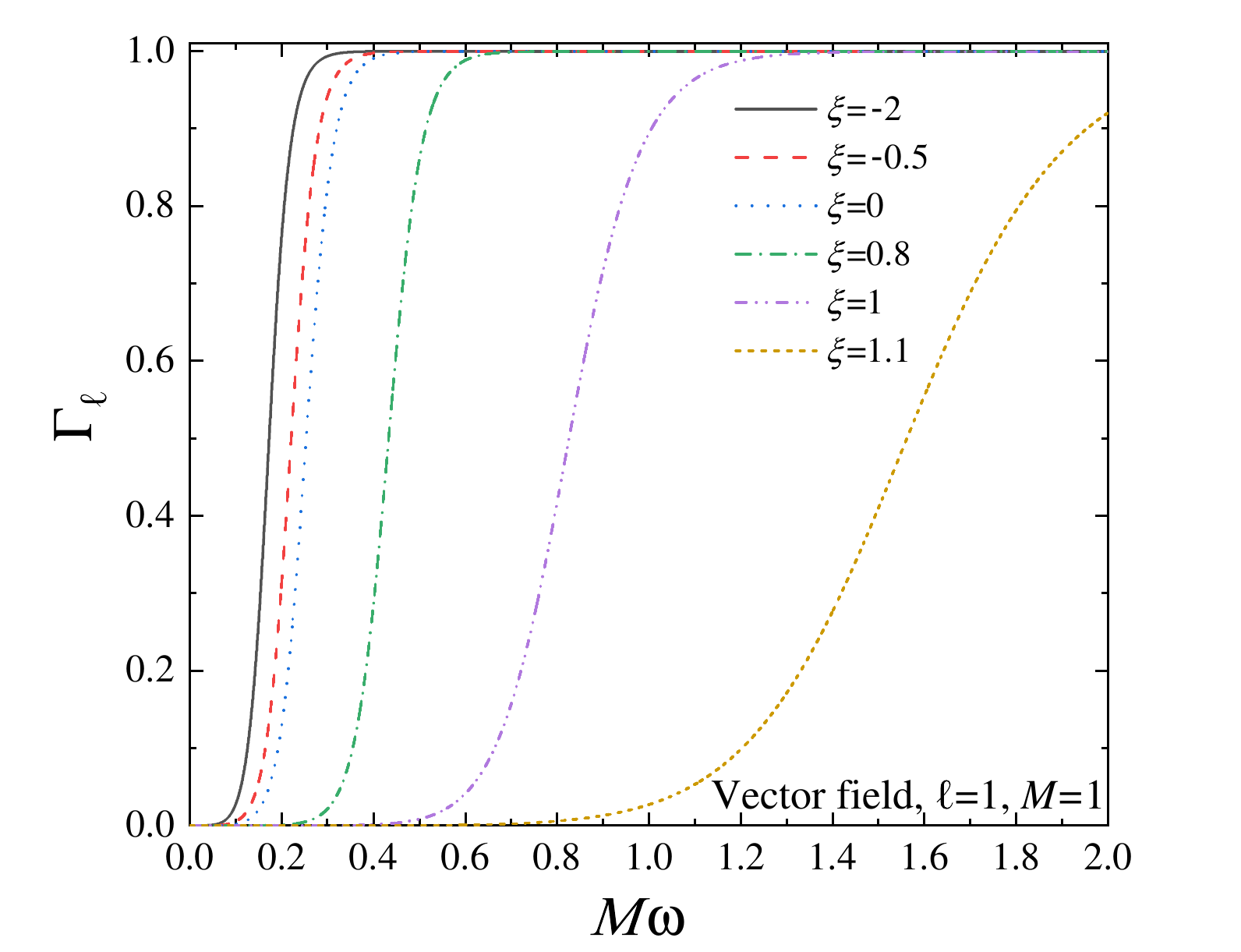}%
    \includegraphics[width=0.33\linewidth]{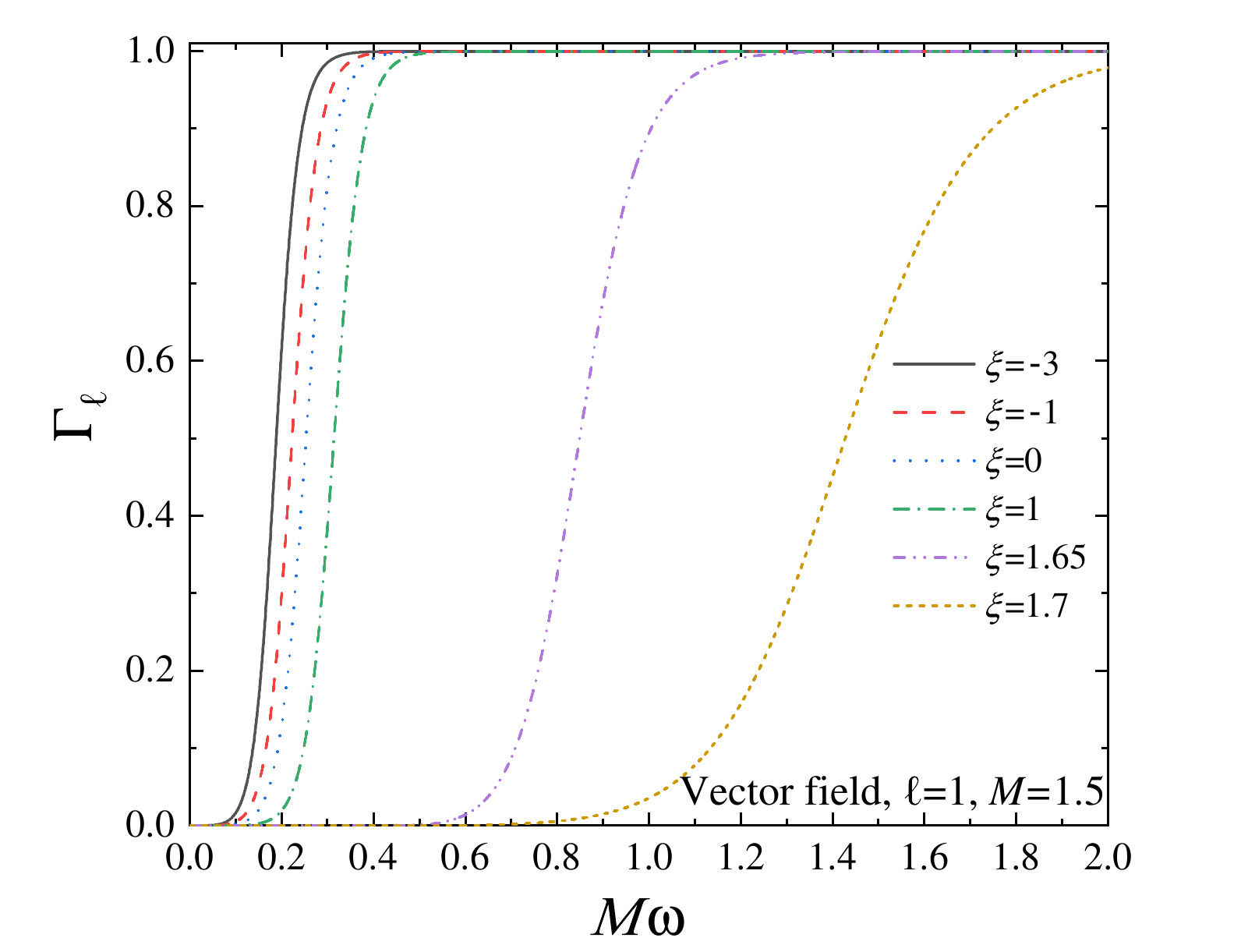}
    \includegraphics[width=0.33\linewidth]{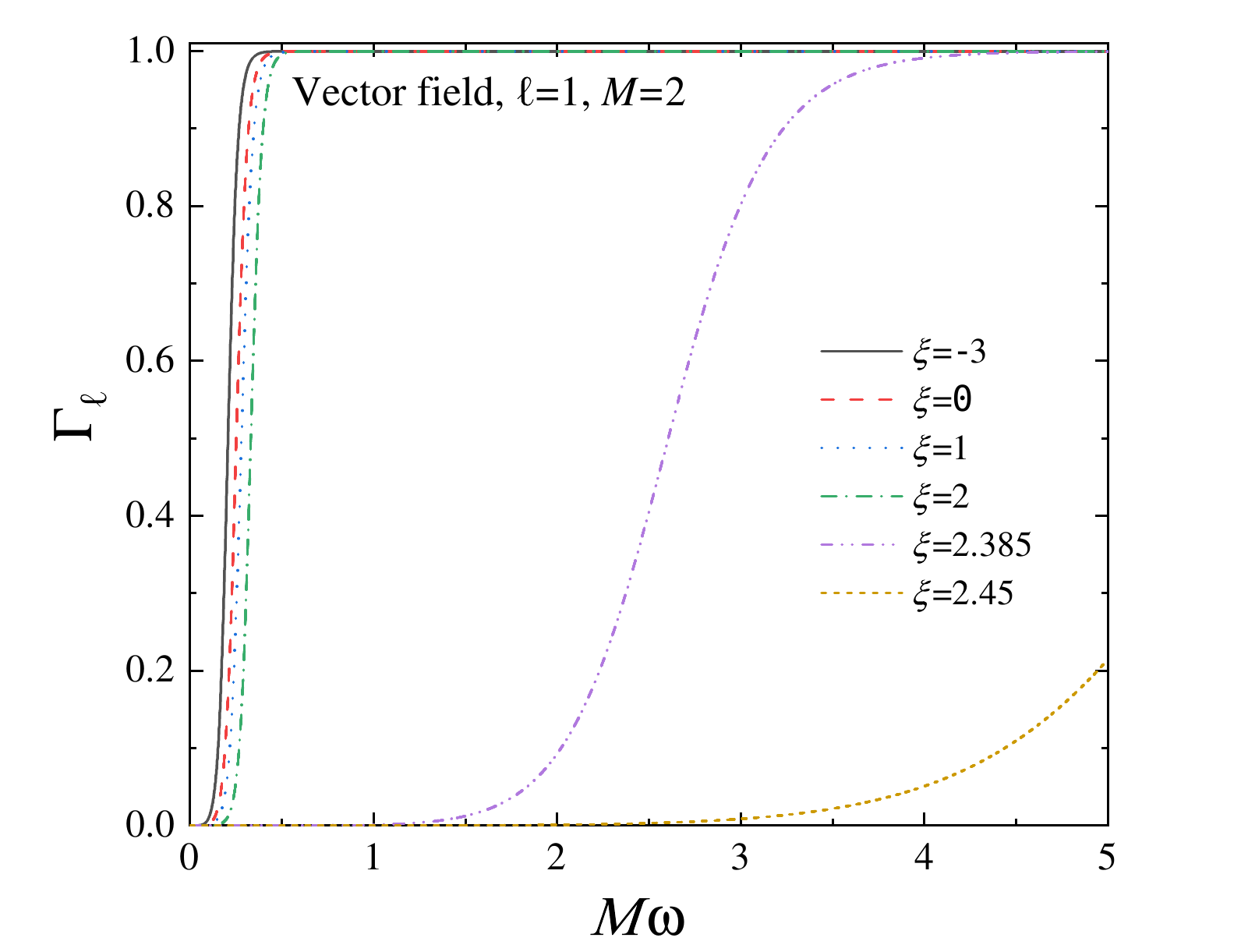}
    \caption{Electromagnetic graybody factors for the lowest allowed multipole $\ell=1$ and representative values of the mass $M$ and hair parameter $\xi$. Increasing or deforming the potential barrier shifts the transmission region and may suppress the low-frequency flux.}%
    \label{fig:GB_vector}
\end{figure*}

\begin{figure*}
    \includegraphics[width=0.33\linewidth]{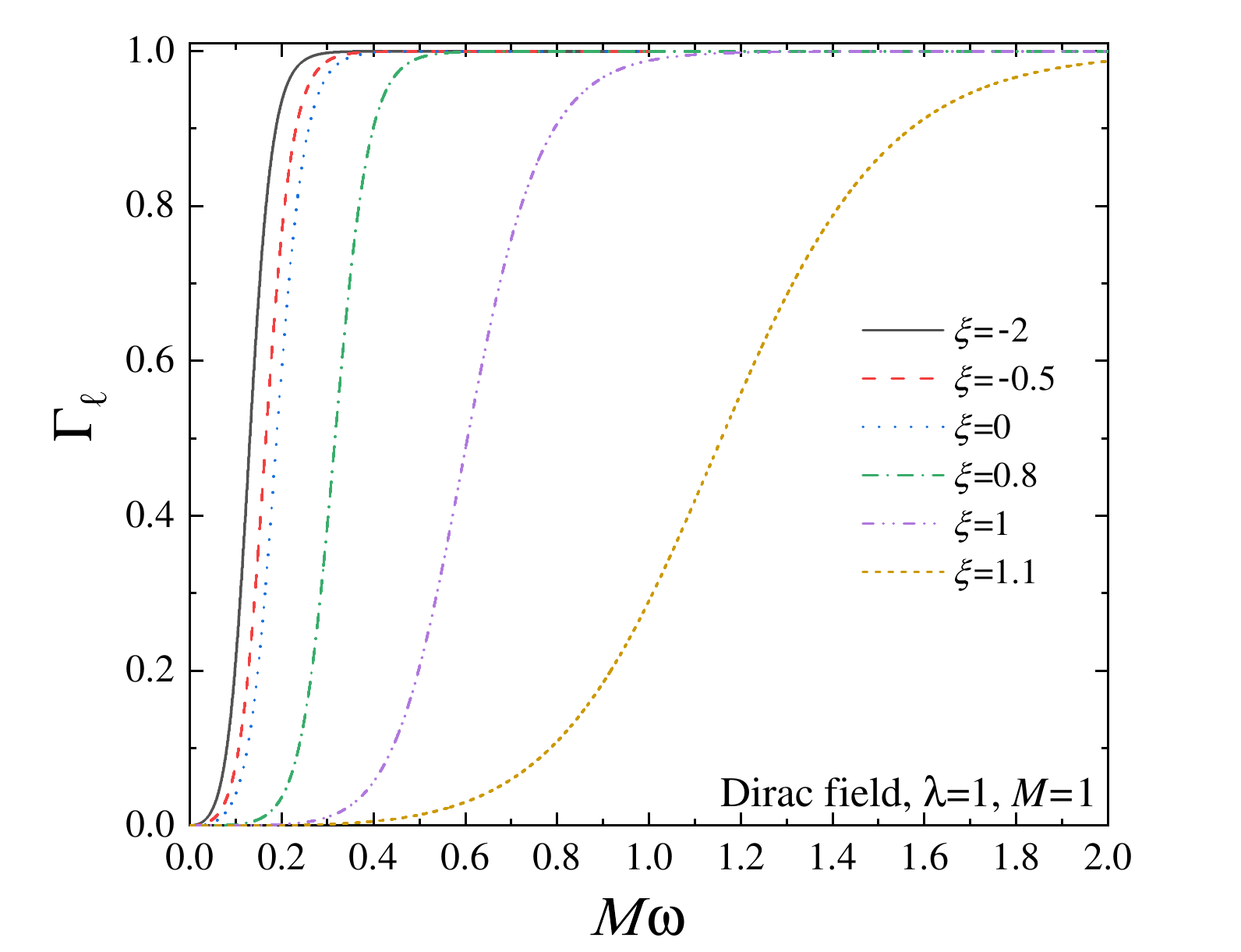}%
    \includegraphics[width=0.33\linewidth]{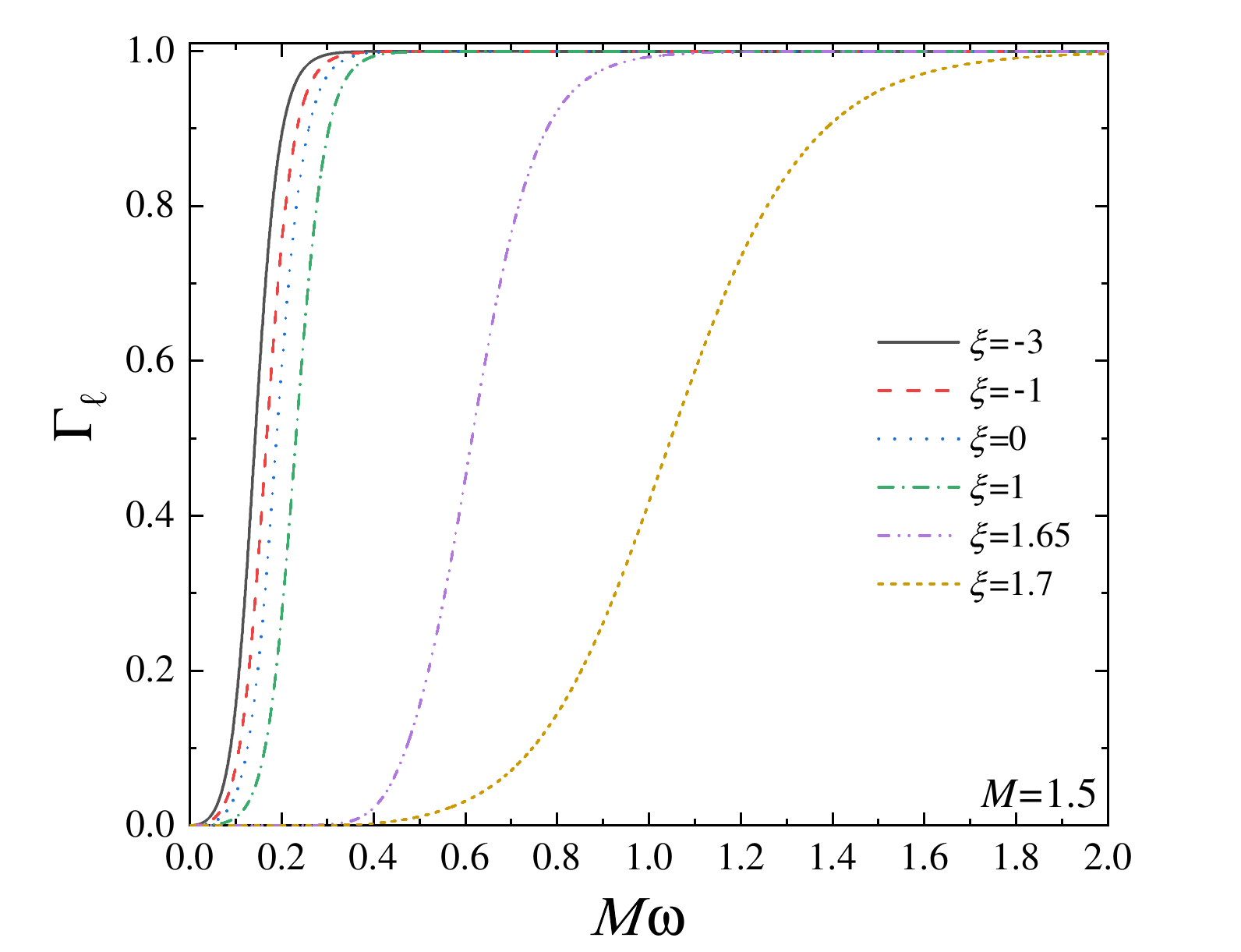}
    \includegraphics[width=0.33\linewidth]{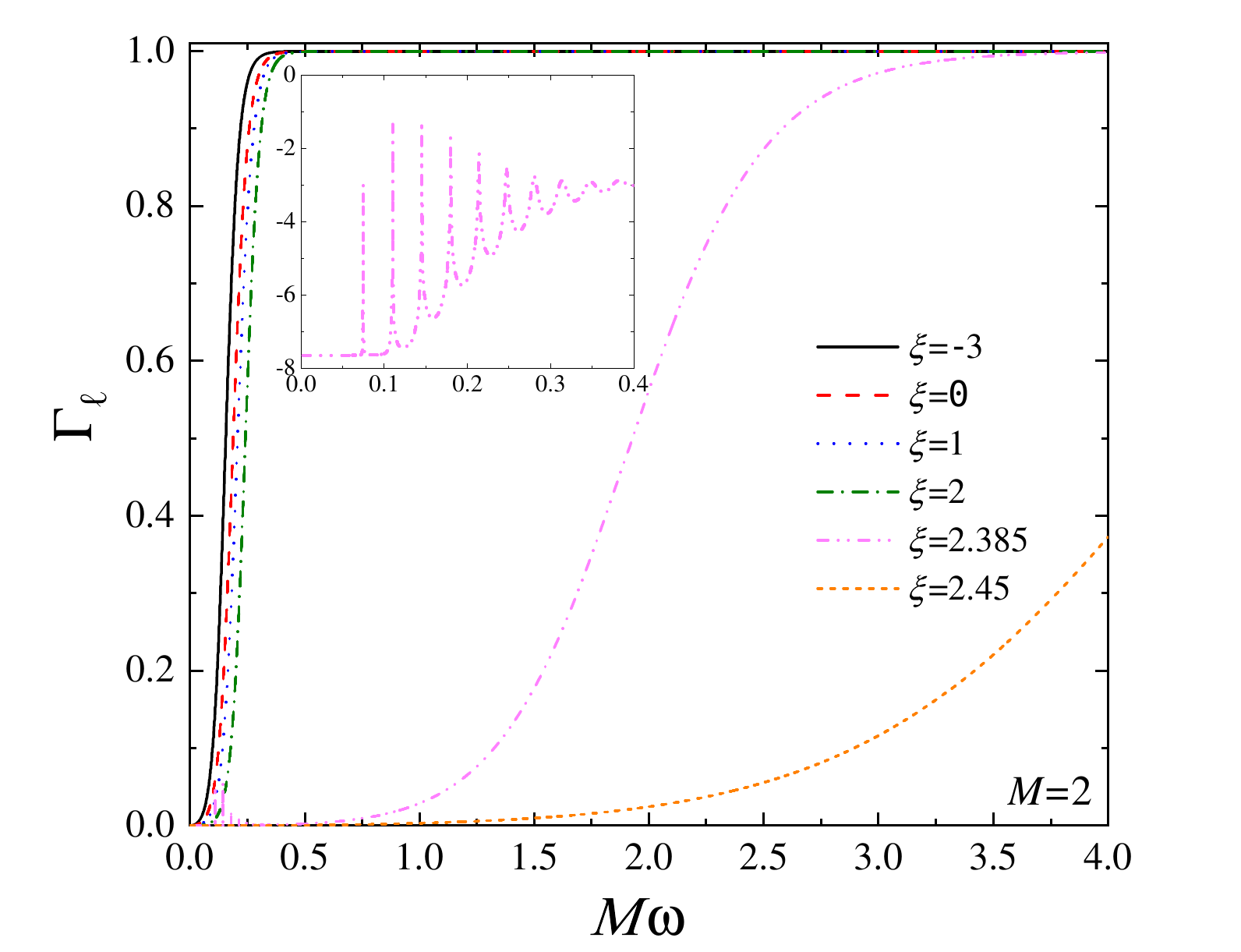}\\

    \caption{Dirac-field graybody factors for the lowest angular mode used in the numerical calculation and representative values of $M$ and $\xi$. The curves display the same qualitative barrier-controlled shift of the transmission window as in the scalar and electromagnetic cases.}
    \label{fig:GB_dirac}
\end{figure*}

\begin{figure*}
    \centering
    \includegraphics[width=0.33\linewidth]{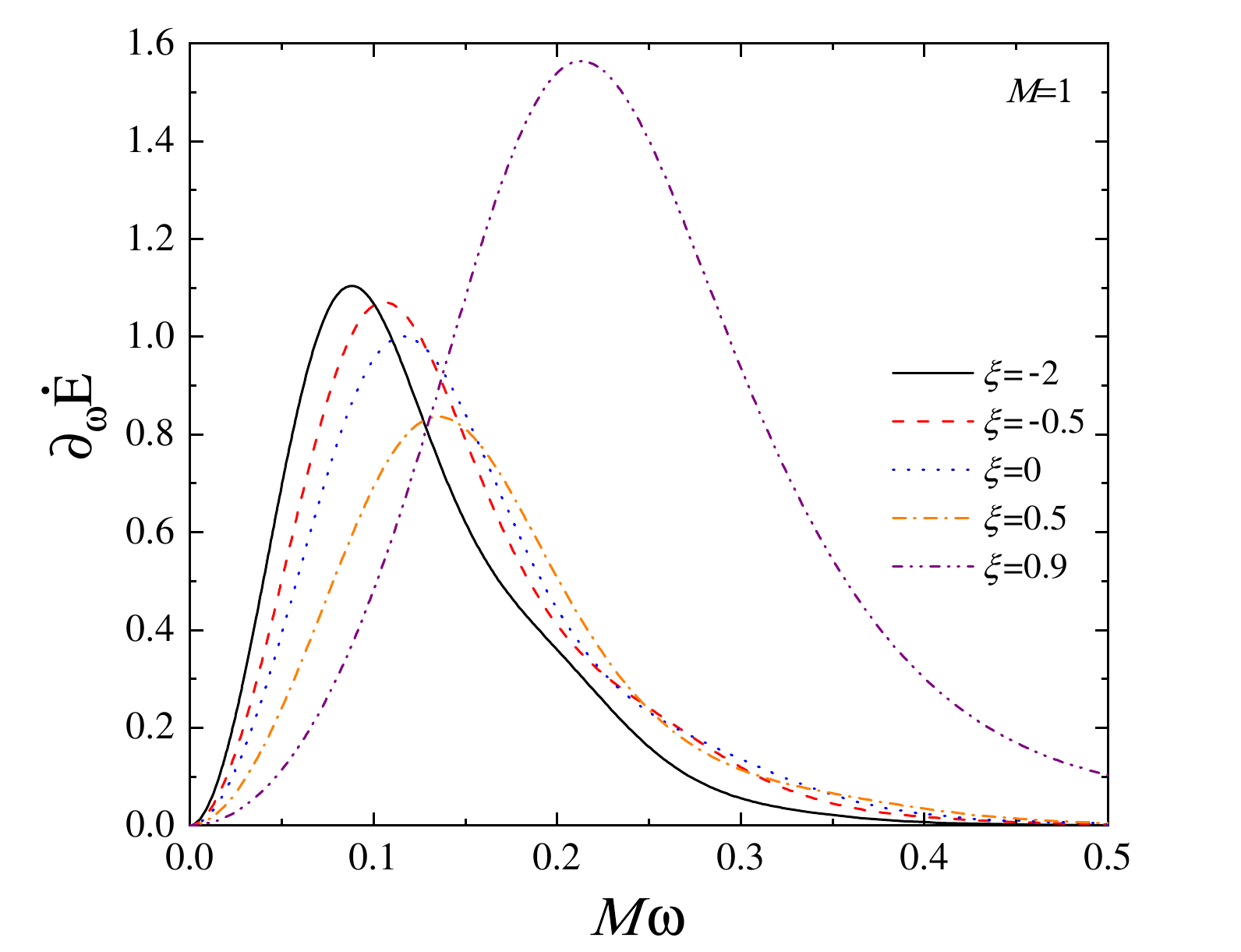}%
    \includegraphics[width=0.33\linewidth]{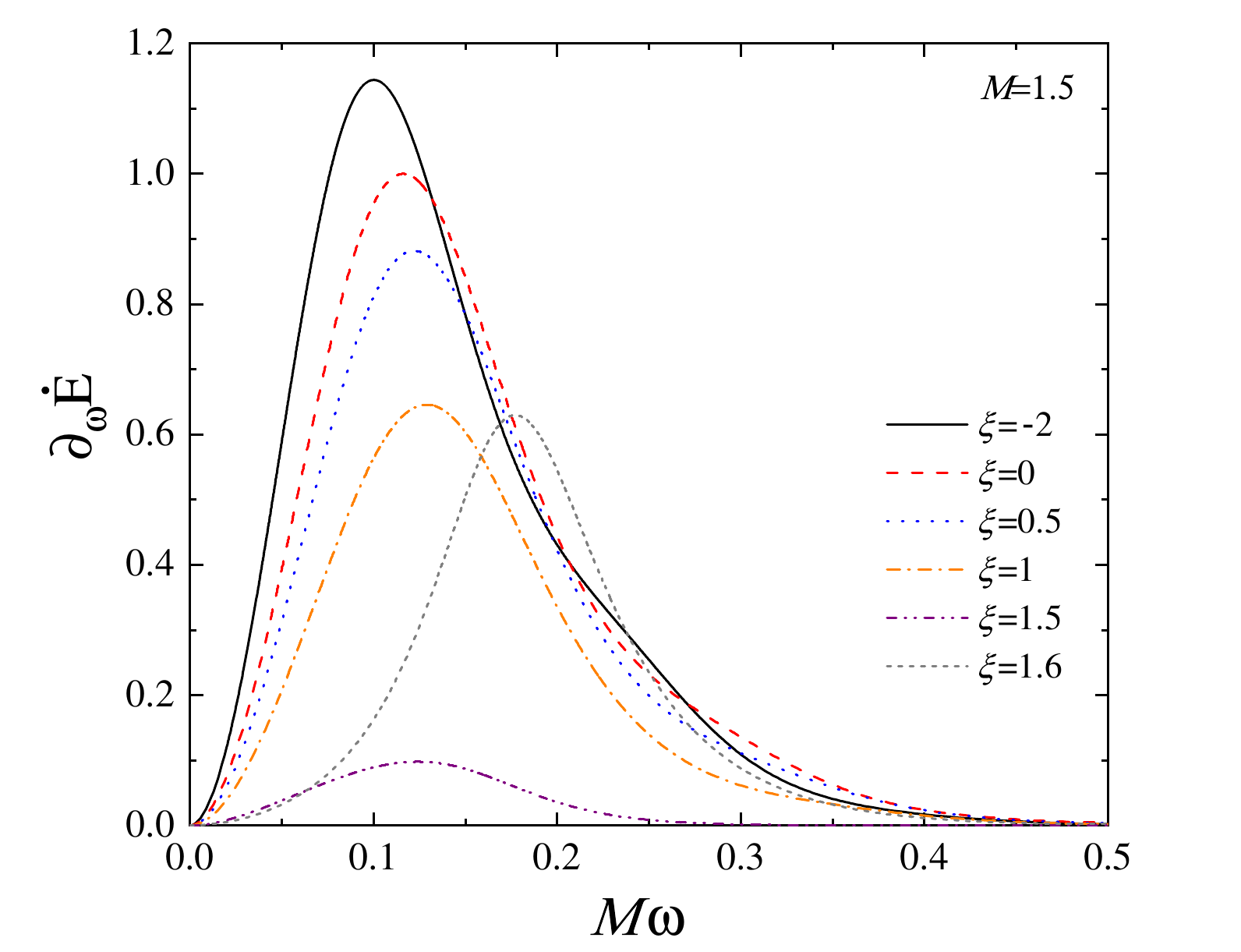}%
        \includegraphics[width=0.33\linewidth]{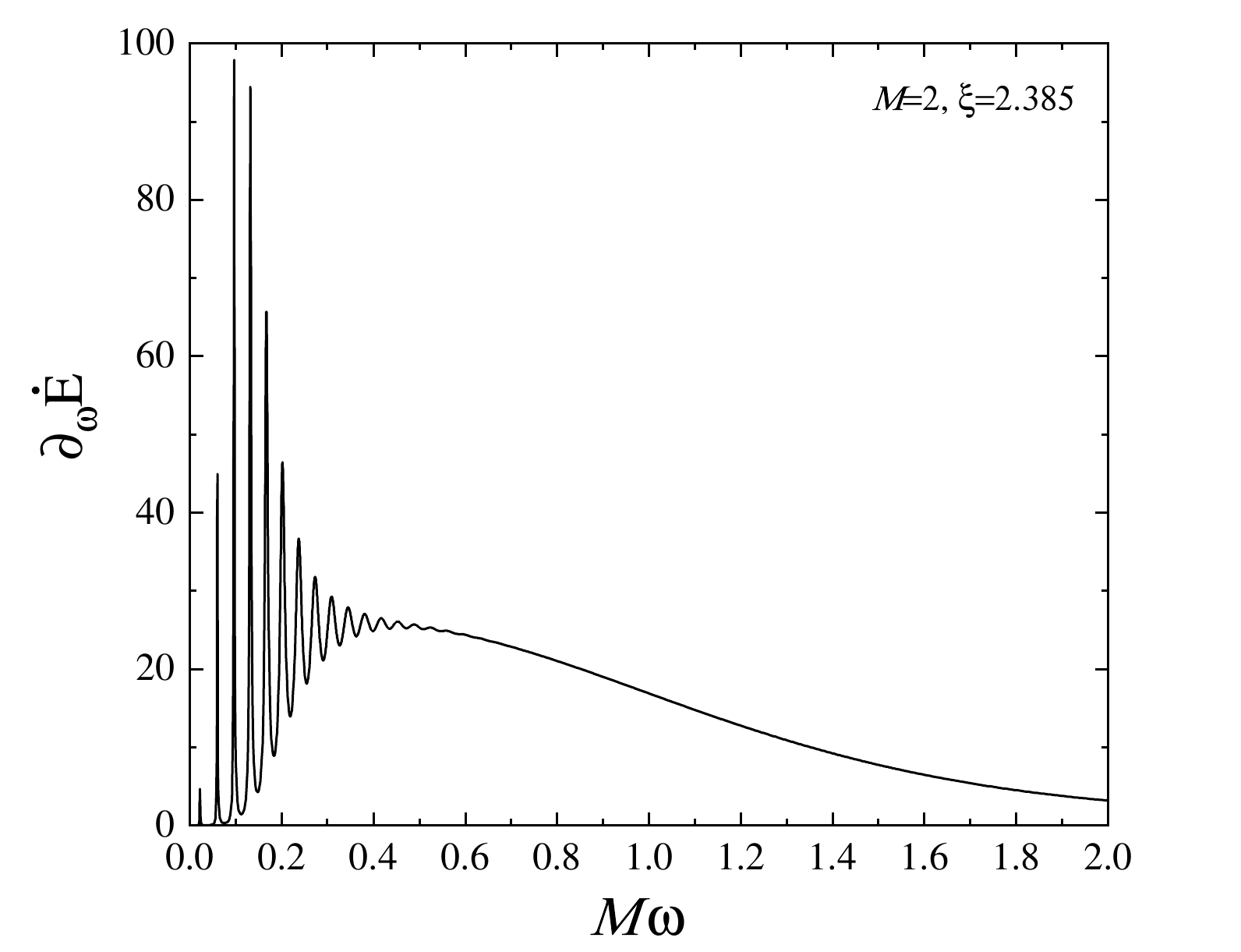}%
    \caption{Differential Hawking-emission spectra for the scalar field for representative values of $M$ and $\xi$. Each curve is normalized by the maximum of the corresponding Schwarzschild case, $\xi=0$, and the first five scalar multipoles are included in the partial-wave sum.}
    \label{fig:Emission_scalar}
\end{figure*}

\begin{figure*}
    \centering
    \includegraphics[width=0.33\linewidth]{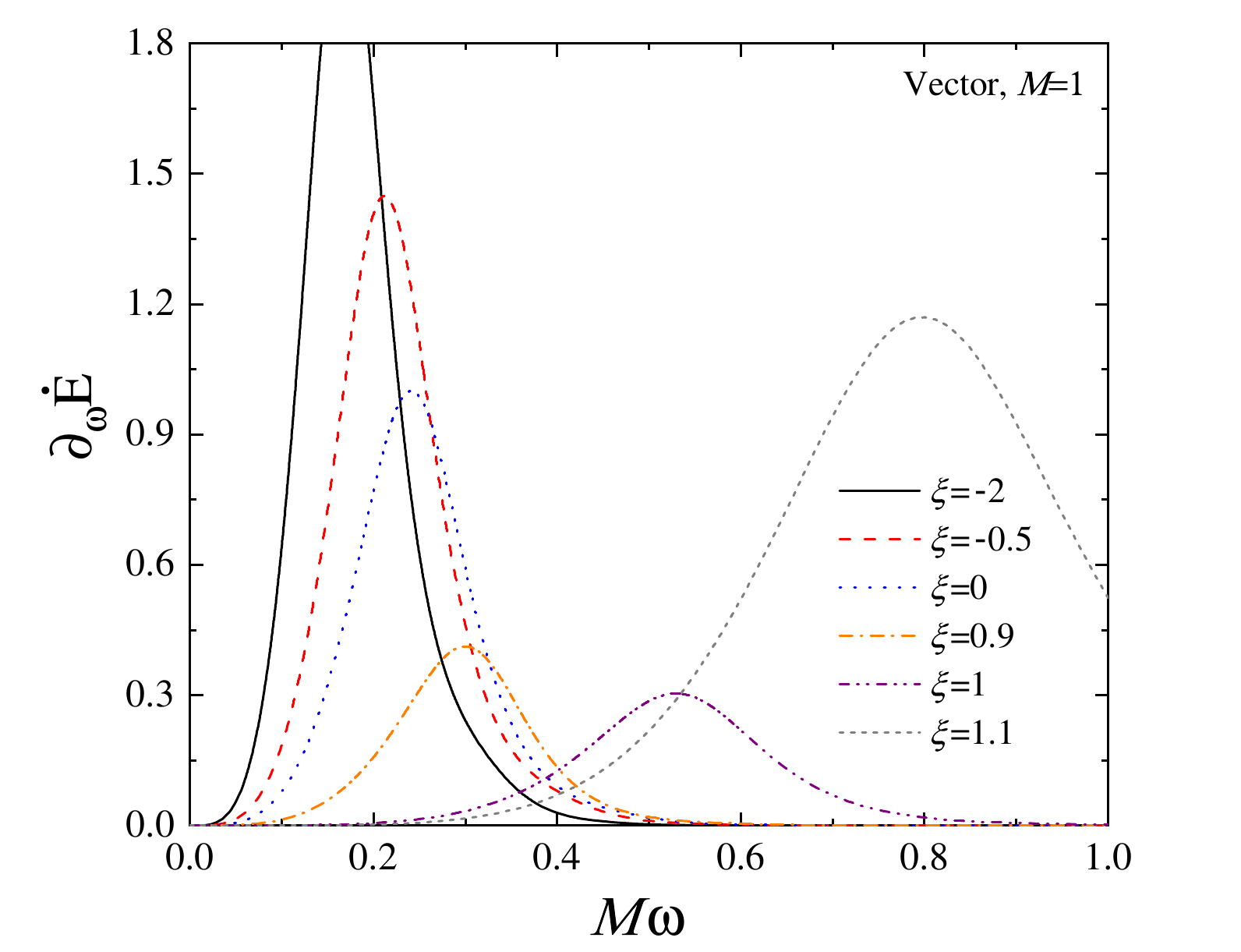}%
    \includegraphics[width=0.33\linewidth]{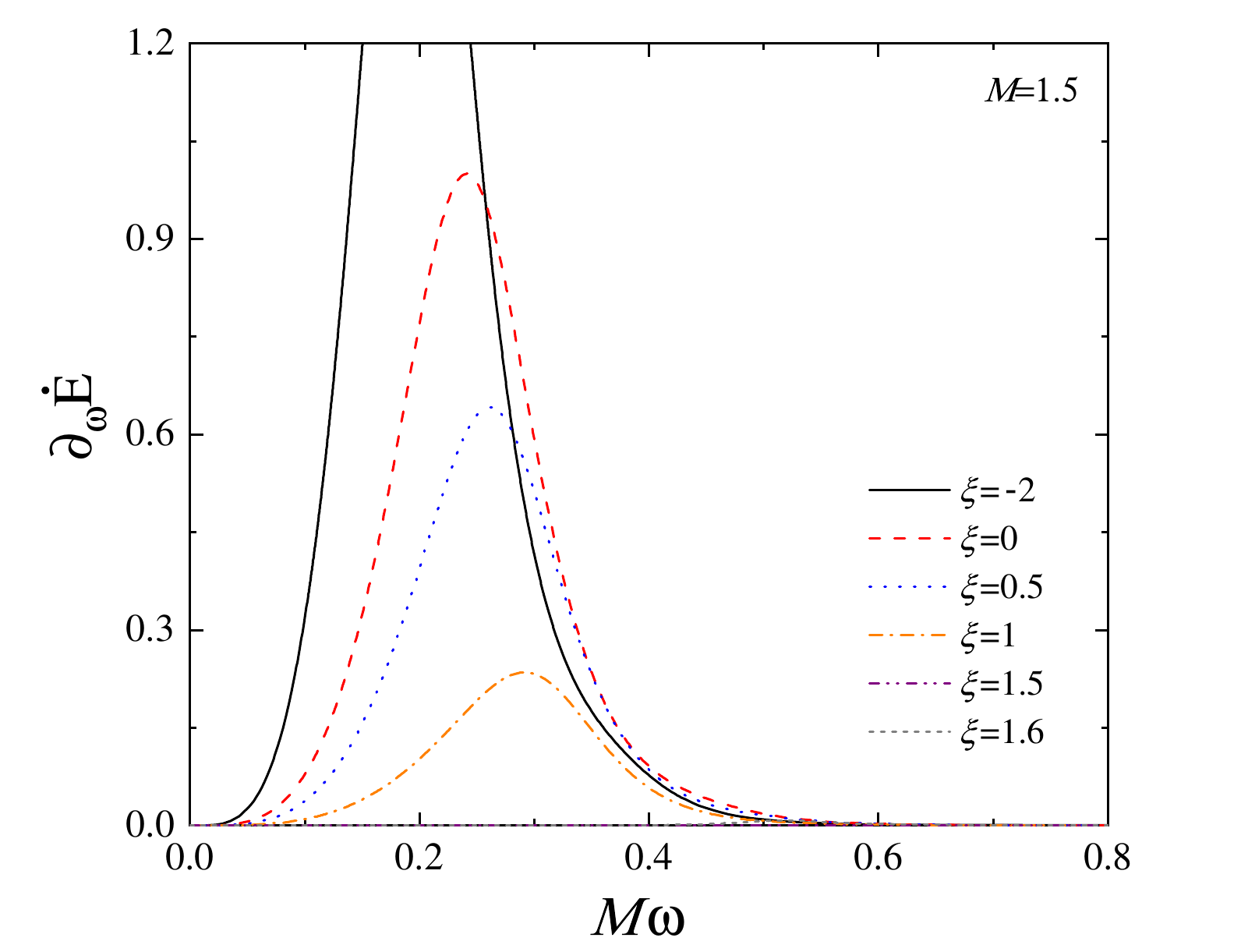}%
        \includegraphics[width=0.33\linewidth]{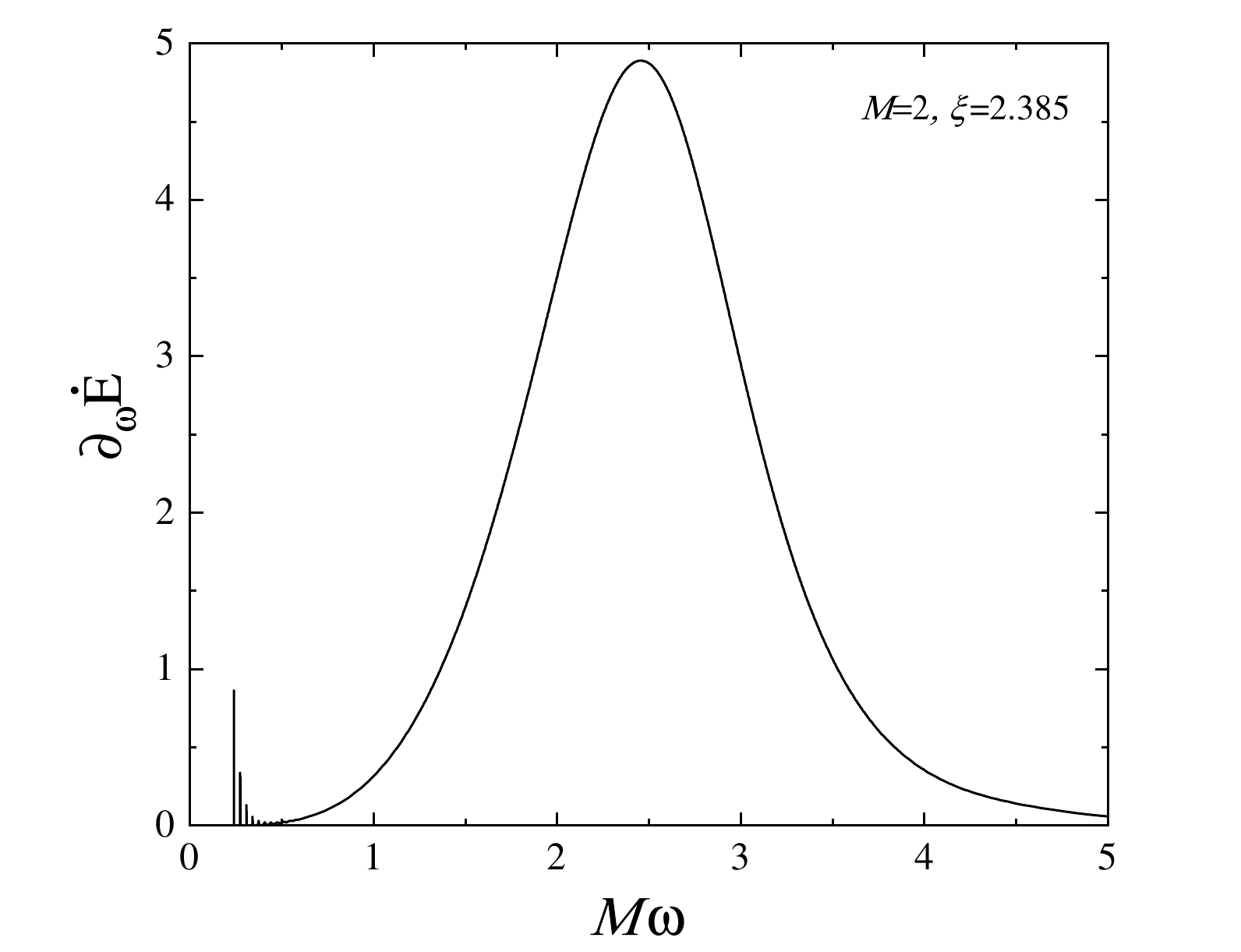}%
    \caption{Differential Hawking-emission spectra for the electromagnetic field for representative values of $M$ and $\xi$. Each curve is normalized by the maximum of the corresponding Schwarzschild case, $\xi=0$, and the first five vector multipoles are included in the partial-wave sum.}
    \label{fig:Emission_vector}
\end{figure*}

\begin{figure*}
    \centering
    \includegraphics[width=0.33\linewidth]{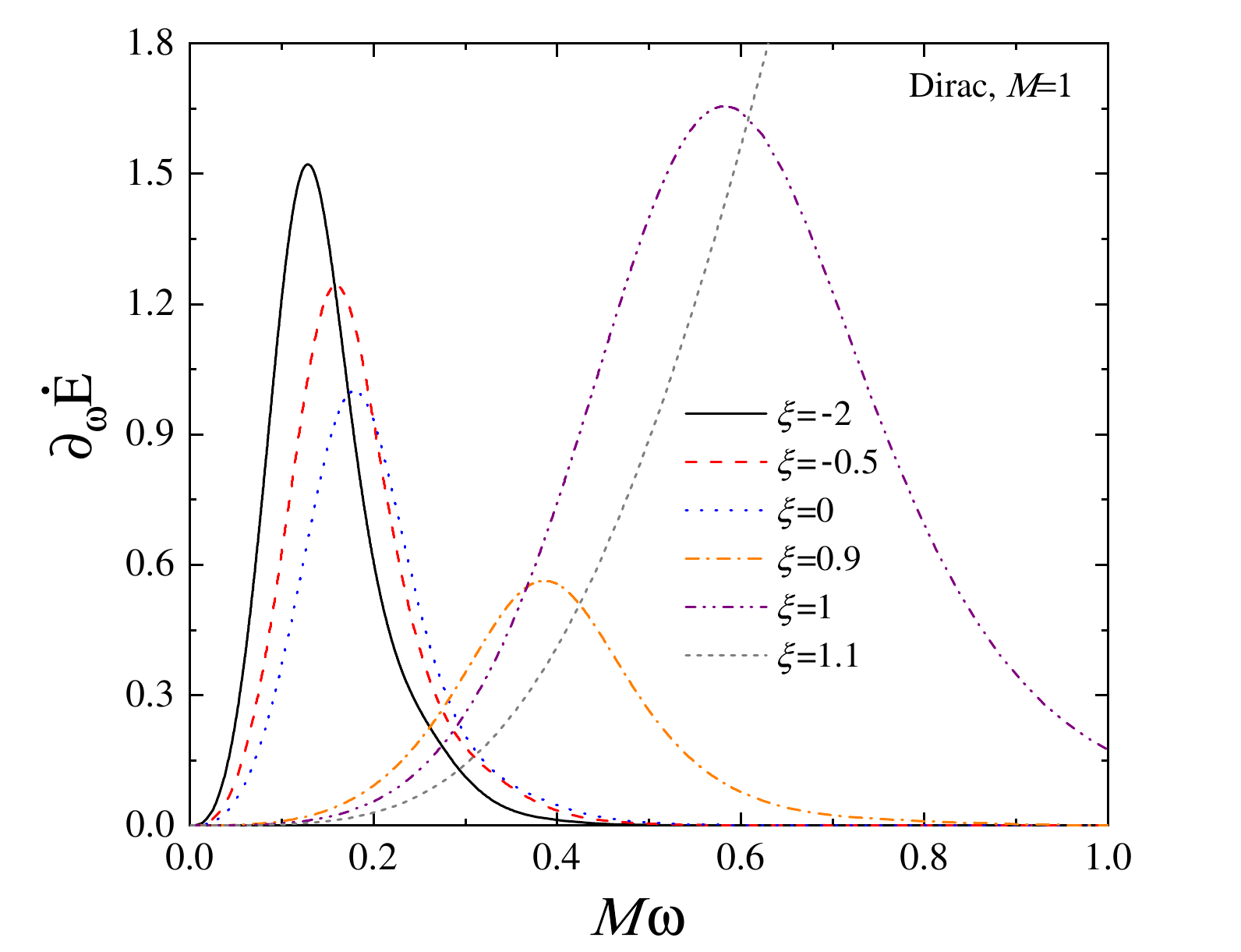}%
    \includegraphics[width=0.33\linewidth]{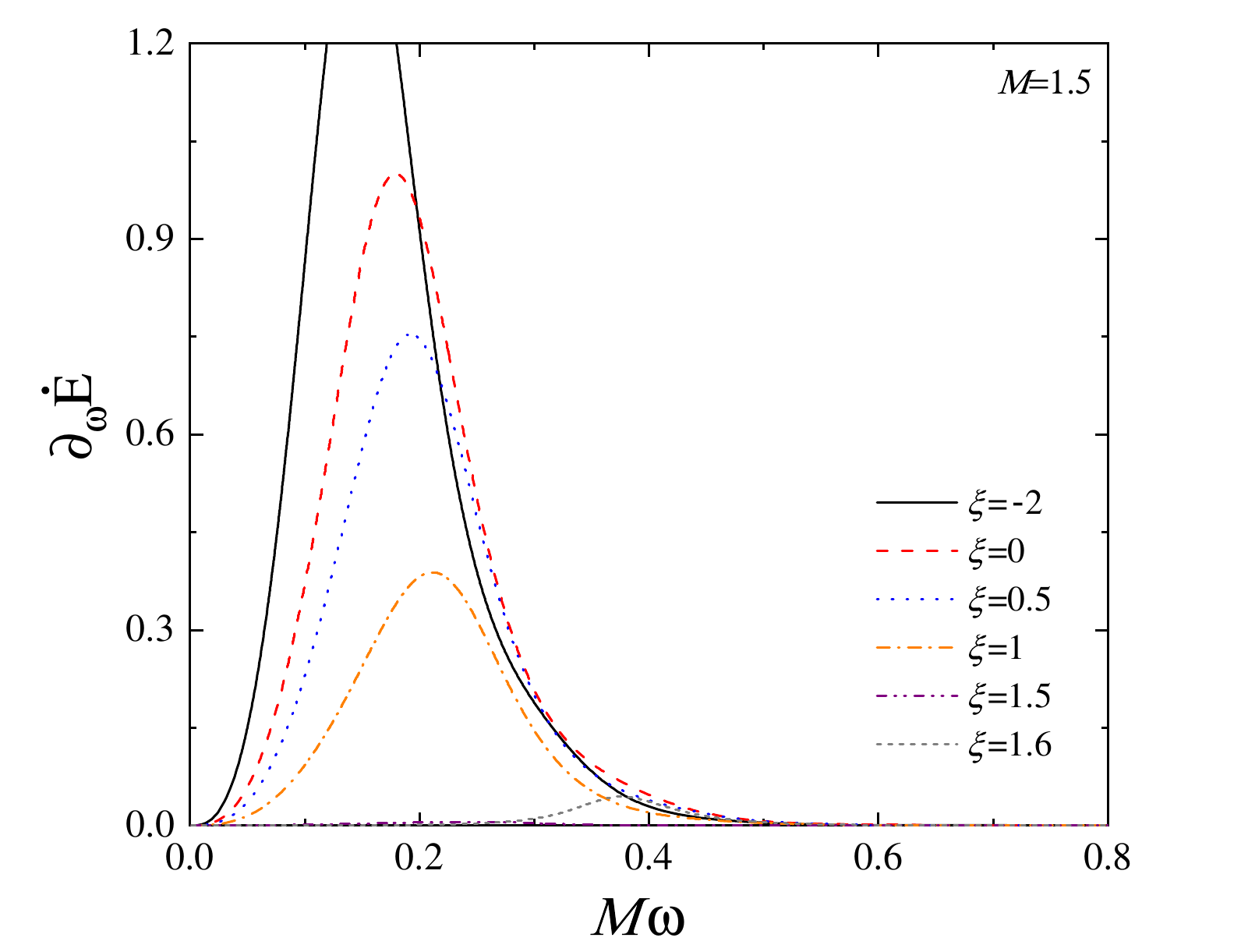}%
        \includegraphics[width=0.33\linewidth]{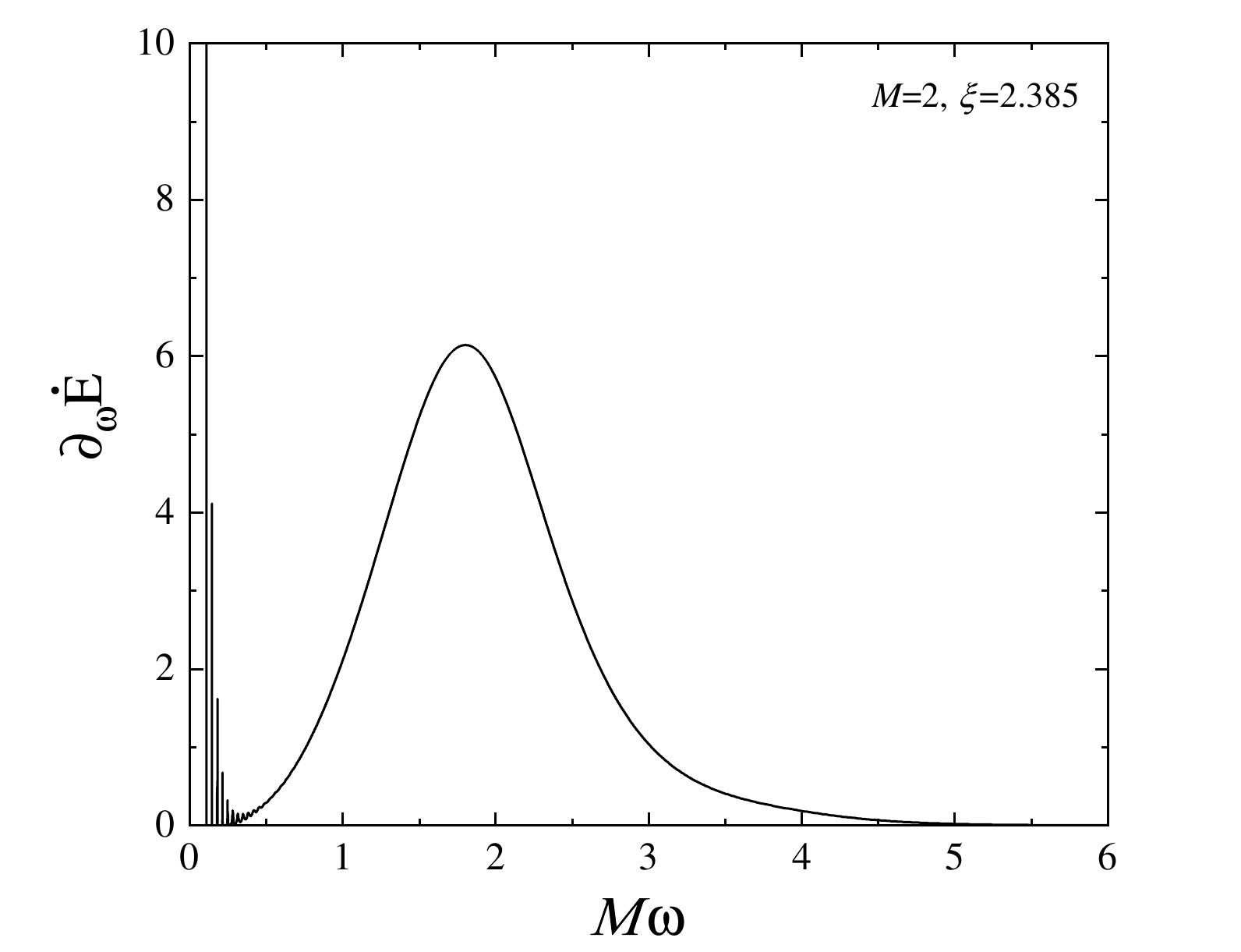}%
    \caption{Differential Hawking-emission spectra for the Dirac field for representative values of $M$ and $\xi$. Each curve is normalized by the maximum of the corresponding Schwarzschild case, $\xi=0$, and the first five angular modes are included in the partial-wave sum.}
    \label{fig:Emission_dirac}
\end{figure*}

\begin{figure*}
    \centering
    \includegraphics[width=0.33\linewidth]{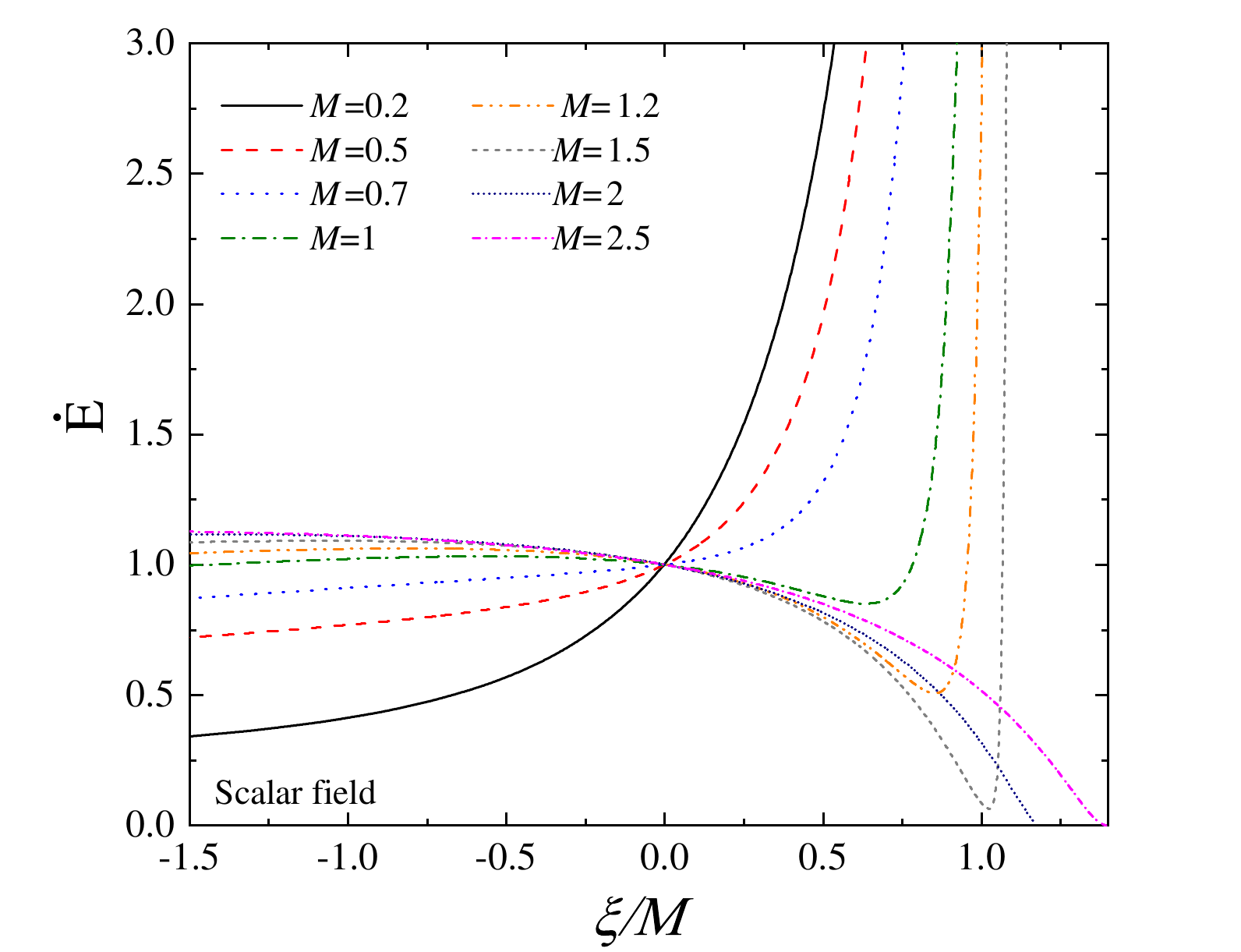}%
    \includegraphics[width=0.33\linewidth]{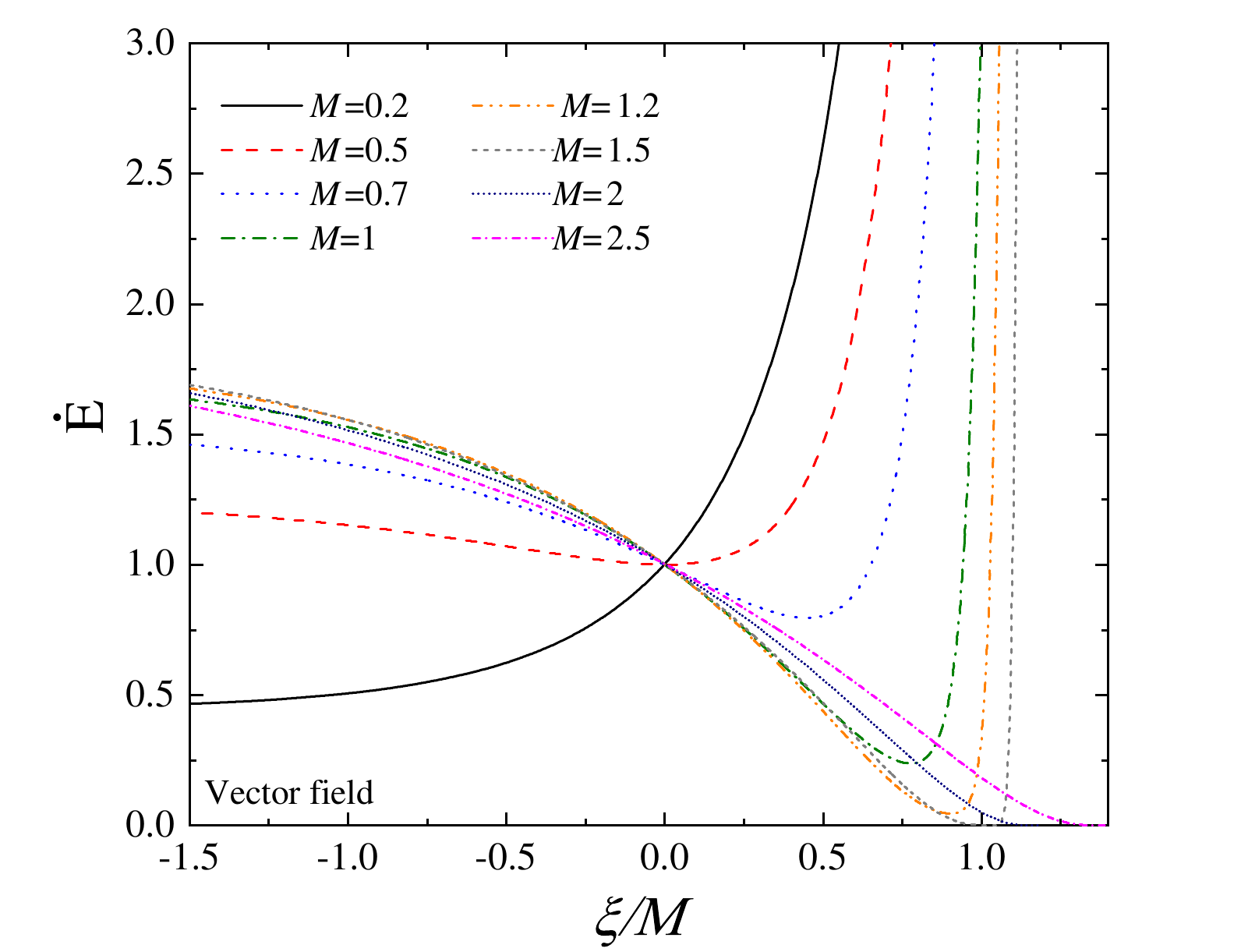}%
        \includegraphics[width=0.33\linewidth]{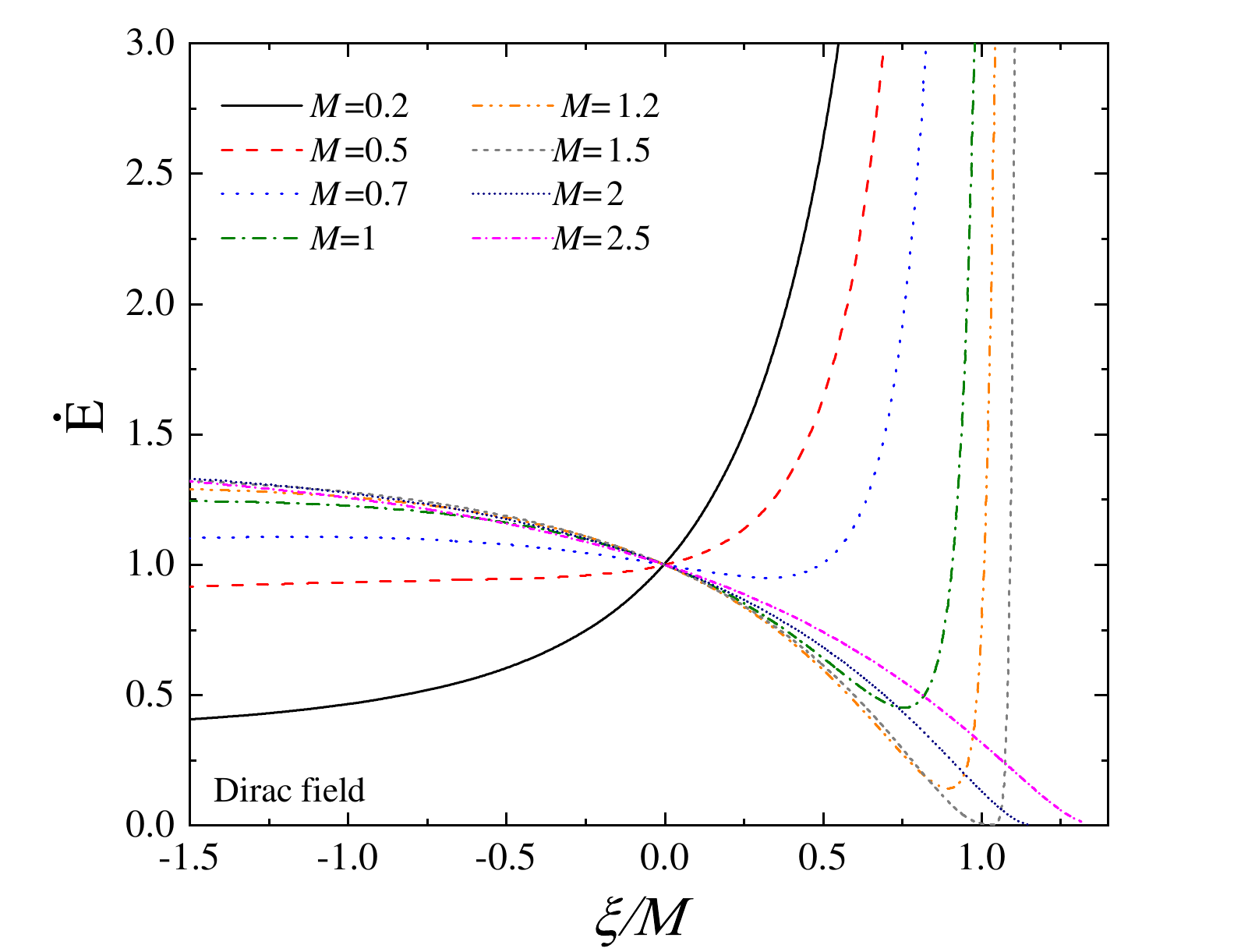}%
    \caption{Total Hawking-emission power as a function of the scalar-hair parameter $\xi$ for scalar, electromagnetic, and Dirac test fields and for several values of the mass $M$. The curves summarize the competition between the Hawking temperature, graybody suppression, and possible horizon-branch transitions.}
    \label{fig:total_em}
\end{figure*}

We numerically calculated the graybody factor and emmission of Hawking radiation for all fields. The results are shown for gray body factor in Figs. \ref{fig:GB_scalar}-\ref{fig:GB_dirac}  and for diffential emission of Hawking radiation in Figs. \ref{fig:Emission_scalar}-\ref{fig:Emission_dirac}, and total emission in Fig. \ref{fig:total_em}.
The scattering and evaporation results should be distinguished conceptually. The graybody factors measure only the transmission through the curvature-induced barrier, whereas the emission spectra also include the thermal occupation factor and the spin-dependent degeneracy of each partial wave.

For example, for  the scalar-field case, see Fig.~\ref{fig:GB_scalar}, the behavior can be understood from the simple WKB approximation: for a single-peak potential, one has $\Gamma_\ell(\omega)\simeq\left[1+\exp\left(2\int_{r_{*1}}^{r_{*2}}\sqrt{V(r_*)-\omega^2}\,dr_*\right)\right]^{-1}$, where $r_{*1}$ and $r_{*2}$ are the turning points. If the barrier becomes higher or wider, the integral increases, and therefore $\Gamma_\ell$ decreases at the same frequency. As a result, the curve $\Gamma_\ell(M\omega)$ is shifted to the right. We have demonstrated that for $\xi<0$, the horizon grows as $r_h\sim\sqrt{-2\xi}$. The effective potential has natural scale $r_h^{-2}$, hence the increase of $r_h$ shifts the transition region of the graybody factor toward smaller frequencies. By contrast, for $\xi>0$, the horizon radius decreases, the scale $r_h^{-2}$ increases, and the transition region is shifted toward larger frequencies. The dependence on the background parameters is nevertheless the same as for the other spins.

For sufficiently large  $\xi>0$, the shrinking of the horizon exposes a region of the effective potential that was previously hidden below the horizon. As a result, the exterior potential develops two maxima separated by a well. The corresponding scattering problem is then of double-barrier type, the wave undergoes multiple reflections inside the well, producing interference and resonant tunneling. Then we have quasibound states due to presence of trapped mode $\omega$ that corresponds to  oscillations, dips, and narrow peaks in the graybody factors.  Near such a resonance the graybody factor can be written in the Breit-Wigner form $\Gamma_\ell(\omega)\simeq \Gamma_{\rm bg}(\omega)+A_n(\,\omega_i)^2/[(\omega-\omega_r)^2+(\omega_i)^2]$. In the left and middle panels, such peaks were not found, although this may be caused by the limited numerical resolution.
The absence of visible resonant peaks in some panels should therefore be understood with caution: very narrow resonances require a frequency step smaller than their width, and broad resonances may be hidden in the smooth background transmission. The location of such structures is correlated with the real parts of the long-lived QNMs of the same double-barrier potential.

The energy emission spectrum normalized by its maximal value at $\xi=0$ is shown in Fig.~\ref{fig:Emission_scalar}. For $\xi<0$, one has $r_h\sim\sqrt{-2\xi}$ and $T_h\sim 1/(2\sqrt{2}\pi\sqrt{-\xi})$. Thus, the thermal scale decreases, and the emission is shifted toward smaller frequencies and suppressed at fixed $M\omega$. For $\xi>0$, the horizon radius decreases, so the potential scale $V\sim r_h^{-2}$ increases and the graybody transition is shifted toward larger frequencies. At the same time, the interplay between $\omega$ and $T_H$ in the thermal factor may either enhance or suppress the emission. The behavior for the other fields is qualitatively similar. In Fig.~\ref{fig:Emission_vector}, the vector spectra show the same competition between the thermal factor and the electromagnetic graybody suppression, while Fig.~\ref{fig:Emission_dirac} confirms that the fermionic thermal factor changes the normalization but not the qualitative dependence on $\xi$. In Fig.~\ref{fig:total_em}, we show the total emission as a function of $\xi$.
The dependence of the total emission on $\xi$ is nonmonotonic and strongly depends on $M$. In the positive-$\xi$ region, the total emission typically develops a minimum due to the competition between the graybody suppression and the increase of the Hawking temperature. For $M_1<M<M_2$, this minimum is accompanied by a discontinuity, because the relevant horizon radius changes discontinuously at the transition point. Thus the jump in $dE/dt$ is directly related to the jump in $r_h$, which changes both the temperature $T_h$ and the graybody factors.
The total flux is dominated by the lowest partial waves. Higher multipoles have larger centrifugal barriers and their contribution is additionally suppressed by the thermal factor at large $\omega/T_h$; this is why the qualitative trends are already visible from the first few $\ell$ modes included in the numerical sum.



\section{Conclusions}

We have studied the propagation of massless scalar, electromagnetic, and Dirac test fields in the asymptotically flat, parity-symmetric beyond-Horndeski black-hole geometry with primary scalar hair. The background is controlled by the mass parameter $M$ and by the hair parameter $\xi$. This two-parameter structure leads to a richer horizon and potential landscape than in the Schwarzschild case, including regions with several horizons, branch changes of the relevant event horizon, and effective potentials with either one or two barriers.

For the quasinormal spectrum, we found that the scalar, vector, and Dirac perturbations show the same main qualitative tendencies. In the negative-$\xi$ region, the horizon radius grows as $r_h\sim\sqrt{-2\xi}$, and the photon-sphere radius grows with the same parametric scale. As a result, both the real oscillation frequency and the damping rate decrease, and in the eikonal regime the QNM frequencies tend to the origin as $\xi\to-\infty$. This behavior follows directly from the geometric scaling of the photon sphere and is consistent with the correspondence between eikonal QNMs and the unstable circular null orbit.

The positive-$\xi$ region is qualitatively more involved. There the deformation of the effective potential may produce additional extrema, mode switching, and eigenvalue repulsion among neighboring overtone branches. These effects are especially visible when the first few overtones are tracked continuously in $\xi$, confirming that overtones are essential for following the spectral flow. For $M>M_1$, the relevant horizon branch can change discontinuously. Then a potential barrier that was previously hidden behind the horizon becomes part of the exterior problem, producing a double-barrier structure, trapped modes, long-lived frequencies, and echoes in the time-domain profile. The resulting spectrum is analogous to that of exotic compact objects and other systems with an effective cavity.

In this sense, the positive-$\xi$ part of the spectrum provides a qualitative example of the ``outburst of overtones'', or the ``sound of the event horizon'', emphasized in~\cite{Konoplya:2022pbc}. The fundamental modes remain comparatively smooth in part of the parameter range, whereas the first few overtones react much more strongly to the near-horizon deformation: they show branch switching, rapid growth of $|{\rm Im}\,\omega|$, and, after the horizon-branch transition, long-lived trapped branches. When the overtone number grows, the rate with which the frequency deviates from its Schwarzschild value increases. This should not be understood as a separate instability, but rather as the expected high sensitivity of overtone branches to the geometry close to the relevant horizon.A similar effect of outburst of overtones has been recently observed for a number of other black-hole configurations, which are deformed from the Schwarzschild/Kerr limit in the near-horizon zone \cite{Konoplya:2023ppx,Konoplya:2023aph,Konoplya:2022iyn,Konoplya:2022hll,Bolokhov:2023ozp}.

We also computed graybody factors and Hawking emission rates for the three test fields. For single-barrier potentials the behavior of the transmission coefficients follows the standard WKB intuition: higher or wider barriers suppress transmission and shift the graybody transition to larger frequencies. In the negative-$\xi$ region the growth of the characteristic length scale shifts the transition to smaller frequencies, whereas for positive $\xi$ the shrinking of the horizon and the increase of the potential scale shift it to larger frequencies. When two barriers are present, interference inside the intermediate well can generate resonant tunneling, leading to oscillations, dips, or narrow peaks in the graybody factors.

The energy-emission spectra reflect the combined effect of the graybody factors and the Hawking temperature. For $\xi<0$, the decreasing temperature suppresses the flux and shifts the spectrum toward lower frequencies. For $\xi>0$, the competition between the increasing temperature and the stronger graybody suppression leads to a nonmonotonic dependence of the total emission on $\xi$. In the parameter range where the horizon branch changes discontinuously, the total emission also develops a discontinuity, because both the temperature and the transmission coefficients change abruptly. The lowest partial waves dominate the total flux, so the qualitative evaporation trends are already captured by the first few multipoles included in the numerical sums.

Our analysis was restricted to test fields on a fixed beyond-Horndeski background. This is sufficient for understanding the dominant qualitative features of wave scattering and Hawking evaporation, but it does not replace a full perturbative treatment of the metric and scalar degrees of freedom of the underlying theory. A natural continuation of this work is therefore to compute the gravitational and scalar-led perturbation spectra of the same black-hole family, to check their stability in the full theory, and to compare their ringdown, echo, and emission signatures with the test-field results obtained here.
\vskip3mm \textit{Acknowledgements.} 

O.~S.~S. is supported in part
by National Science Foundation grant PHY-2110466 and
the Tufts Scholar at Risk Program.
\clearpage
\bibliographystyle{apsrev4-2}
\bibliography{refs}

\end{document}